\documentclass[]{aa}  
\usepackage{graphicx}
\usepackage{txfonts}
\usepackage{placeins}

\begin{document}

   \title{Slow and steady does the trick: Slow outflows enhance the fragmentation of molecular clouds}
    \subtitle{}
   \author{M.~Laužikas
          \orcid{0000-0002-9395-9207}
          \inst{1}
          \and
          K.~Zubovas
          \orcid{0000-0002-9656-6281}
          \inst{1, 2}
          }

   \institute{Center for Physical Sciences and Technology, Saulėtekio al. 3, Vilnius LT-10257, Lithuania\\
         \email{martynas.lauzikas@ftmc.lt}
         \and
         Astronomical Observatory, Vilnius University, Saulėtekio al. 3, Vilnius LT-10257, Lithuania
         }

   \date{Received; accepted}

  \abstract
   {Most massive galaxies host a supermassive black hole at their centre. Matter accretion creates an active galactic nucleus (AGN), forming a relativistic particle wind. The wind heats and pushes the interstellar medium, producing galactic-wide outflows. Fast outflows remove the gas from galaxies and quench star formation, and while slower ($v<500$~km~s$^{-1}$) outflows are ubiquitous, their effect is less clear but can be both positive and negative.}
   {We wish to understand the conditions required for positive feedback. We investigated the effect that slow and warm-hot outflows have on the dense gas clouds in the host galaxy. We aim to constrain the region of outflow and cloud parameter space, if any, where the passage of the outflow enhances star formation.}
   {We used numerical simulations of virtual `wind tunnels' to investigate the interaction of isolated turbulent spherical clouds ($10^{3;4;5}$~M$_{\sun}$) with slow outflows ($10$~km~s$^{-1} \leq v_{\rm out} \leq 400$~km~s$^{-1}$) spanning a wide range of temperatures ($10^{4;5;6}$~K). We modelled 57 systems in total.}
   {We find that warm outflows compress the clouds and enhance gas fragmentation at velocities ${\leq}200$~km~s$^{-1}$, while hot ($T_{\rm out} = 10^6$~K) outflows increase fragmentation rates even at moderate velocities of $400$~km~s$^{-1}$. Cloud acceleration, on the other hand, is typically inefficient, with dense gas only attaining velocities of ${<}0.1 v_{\rm out}$.}
   {We suggest three primary scenarios where positive feedback on star formation is viable: stationary cloud compression by slow outflows in low-powered AGN, sporadic enhancement in shear flow layers formed by luminous AGN, and self-compression in fragmenting AGN-driven outflows. We also consider other potential scenarios where suitable conditions arise, such as compression of galaxy discs and supernova explosions. Our results are consistent with current observational constraints and with previous works investigating triggered star formation in these disparate domains.}


   \maketitle
%
\nolinenumbers
\section{Introduction}\label{sec:intro}

It is by now well established that centres of massive galaxies are hosts to super-massive black holes (SMBHs). Accretion onto an SMBH creates an active galactic nucleus (AGN) \citep{Ho:2008:} and often launches a quasi-relativistic particle wind. The wind interacts with the diffuse interstellar medium (ISM), forming shocks and discontinuities and driving massive large-scale outflows \citep{KingPounds:2015:}. Over the past few decades, outflows have been detected in galaxies at various stages of evolution \citep[for a comprehensive review, see][]{VeilleuxMaiolino:2020:, LahaReynolds:2021:}, suggesting numerous scenarios for feedback on the surrounding ISM. It has been argued that outflows play a major role in gas transport, regulating star formation and therefore establishing the observed SMBH-galaxy scaling relations, such as $M\text{--}\sigma$ \citep{FerrareseMerritt:2000:, ZubovasKing:2019:}.

Massive outflows, common in active galaxies, are typically found within $10$~kpc from the AGN, with velocities of $100 \text{ to } 1000$~km~s$^{-1}$ \citep{VeilleuxMaiolino:2020:, LahaReynolds:2021:}. These values match the predictions of the semi-analytic energy-driven outflow model derived by \citet{ZubovasKing:2016:}. It has been suggested that such outflows have a negative feedback effect over the long term (${\sim}100$~Myr), as they remove the gas of a galaxy, thus stifling star formation, further inhibiting SMBH growth, and leaving galaxies `red and dead' \citep{SchawinskiUrry:2014:}. However, observations have also indicated simultaneous negative and positive feedback at the boundary of outflow-blown cavities \citep{CresciMainieri:2015:}. Moreover, local galaxies have shown a positive correlation between AGNs luminosity and star formation rates in circumnuclear regions \citep{Dahmer-HahnRiffel:2022:}, raising the possibility that an AGN can facilitate star formation. The predictions of semi-analytic models are in partial agreement with observations -- outflows can produce external pressure, compressing the cold gas and thus enhancing star formation \citep{Silk:2013:, ZubovasNayakshin:2013:b}. Positive feedback is also possible within the outflow itself as the gas cools down \citep{ZubovasKing:2014:, ThompsonQuataert:2016:}. This prediction has been confirmed by observations \citep{MaiolinoRussell:2017:, GallagherMaiolino:2019:}. These pieces of evidence indicate a diverse effect of AGN activity on the surrounding medium. Nevertheless, the spatial and temporal scales for AGN-enhanced star formation remain uncertain. The uncertainty is not surprising, however, as semi-analytical models ignore various complications found in real galaxies, such as the presence of dense gas structures. The structures alter the flow dynamics and induce mixing, leading to a reduction of typical outflow velocities \citep{FluetschMaiolino:2021:}. Such slow outflows can enhance star formation locally, as seen in numeric 3D simulations, while faster outflows provide a negative global effect \citep[e.g.][]{BieriDubois:2016:, Mercedes-FelizAngles-Alcazar:2023:}. However, the aforementioned simulations rely on sub-resolution methods to account for the small-scale effects of the mixing between the molecular cloud and the outflow. The detailed results of such simulations depend on the selected sub-resolution prescription \citep{WursterThacker:2013:, ValentiniMurante:2017:}, raising the question of whether such methods are sufficiently robust for realistic environments.

Individual cloud-outflow interactions have been studied extensively both analytically and with numerical simulations. Early studies found that supersonic outflows disperse molecular clouds \citep{KleinMcKee:1994:}. However, they exclude several crucial physical processes, such as radiative cooling, turbulence, and self-gravity. A similar approach was taken by \citet{CooperBicknell:2009:}, who included density gradients in the clouds and cooling but did not account for turbulent velocities. Currently, there is little doubt that fast galactic outflows, with velocities $v_{\rm out} \ga 1000$~km~s$^{-1}$, disperse molecular clouds and quench star formation \citep{PittardHartquist:2010:, HopkinsElvis:2010:}. On the other hand, slower outflows can compress the clouds without dispersing them, resulting in positive feedback. For example, \citet{ZubovasSabulis:2014:} and \citet{DuganGaibler:2017:} have independently showed, using simulations with different cloud density profiles and turbulent field structures, that outflows with a velocity of $300$~km~s$^{-1}$ or lower can enhance star formation. However, despite the plethora of studies (conveniently summarised in \citet[][Table 1]{Banda-BarraganParkin:2016:}, \citet[][Table 1]{DuganGaibler:2017:}), there is no clear answer regarding the parameters that determine whether the clouds are destroyed or their fragmentation is enhanced and how rapidly the two outcomes are achieved.

In this paper, we aim to identify and constrain the outflow properties required for positive AGN feedback. We focus primarily on slower (i.e. with radial velocity or velocity difference between cold and warm phases ${<}500$~km~s$^{-1}$) outflows with a radial distance from the AGN on the order of kiloparsecs. We used an enhanced version of the public SPH/N-body code {\sc Gadget 4} supplemented with a radiative cooling prescription to model the interaction between individual turbulent molecular clouds and warm-hot galactic outflows. We simulated $57$ systems -- virtual `wind tunnels' -- and we aim to identify a region, or regions, in the parameter space of cloud mass, outflow velocity, and temperature where star formation is enhanced or quenched. We considered three values of molecular cloud mass, $M_{\rm cl} = 10^3, 10^4, 10^5$~M$_{\sun}$; six values of outflow velocity $v_{\rm out} \leq 400$~km~s$^{-1}$; and three values of outflow temperature, $T_{\rm out} = 10^4, 10^5, 10^6$~K. We find that $10^6$~K outflows compress the clouds and provide positive feedback throughout the investigated velocity range, while at lower outflow temperatures, the velocity threshold value for positive feedback is reduced to ${\la}200$~km~s$^{-1}$. We propose that star formation-enhancing regions are likely to develop in gas-rich galaxies within the outflows and their surroundings. We do not exclude outflows faster than those considered in this work, as they can also sporadically form bursts of star formation where shear flow develops. Notably as well, the outflow itself can fragment and thus form the stars within.

The paper is structured as follows. In Sect.~\ref{sec:Theoretical background}, we introduce the theoretical background and relevant evolutionary timescales. In Sect.~\ref{sec:Numeric methods}, we describe the numerical methods used and present the results in Sect.~\ref{sec:Results}. We discuss the applicability of our results, peculiarities of outflow-enhanced star formation, and caveats of our models in Sect.~\ref{sec:Discussion}, and we conclude in Sect.~\ref{sec:Conclusions}.

\section{Theoretical background}
\label{sec:Theoretical background}

\subsection{AGN-driven outflows}
\begin{figure}
     \centering
        \resizebox{0.9\hsize}{!}{\includegraphics[]{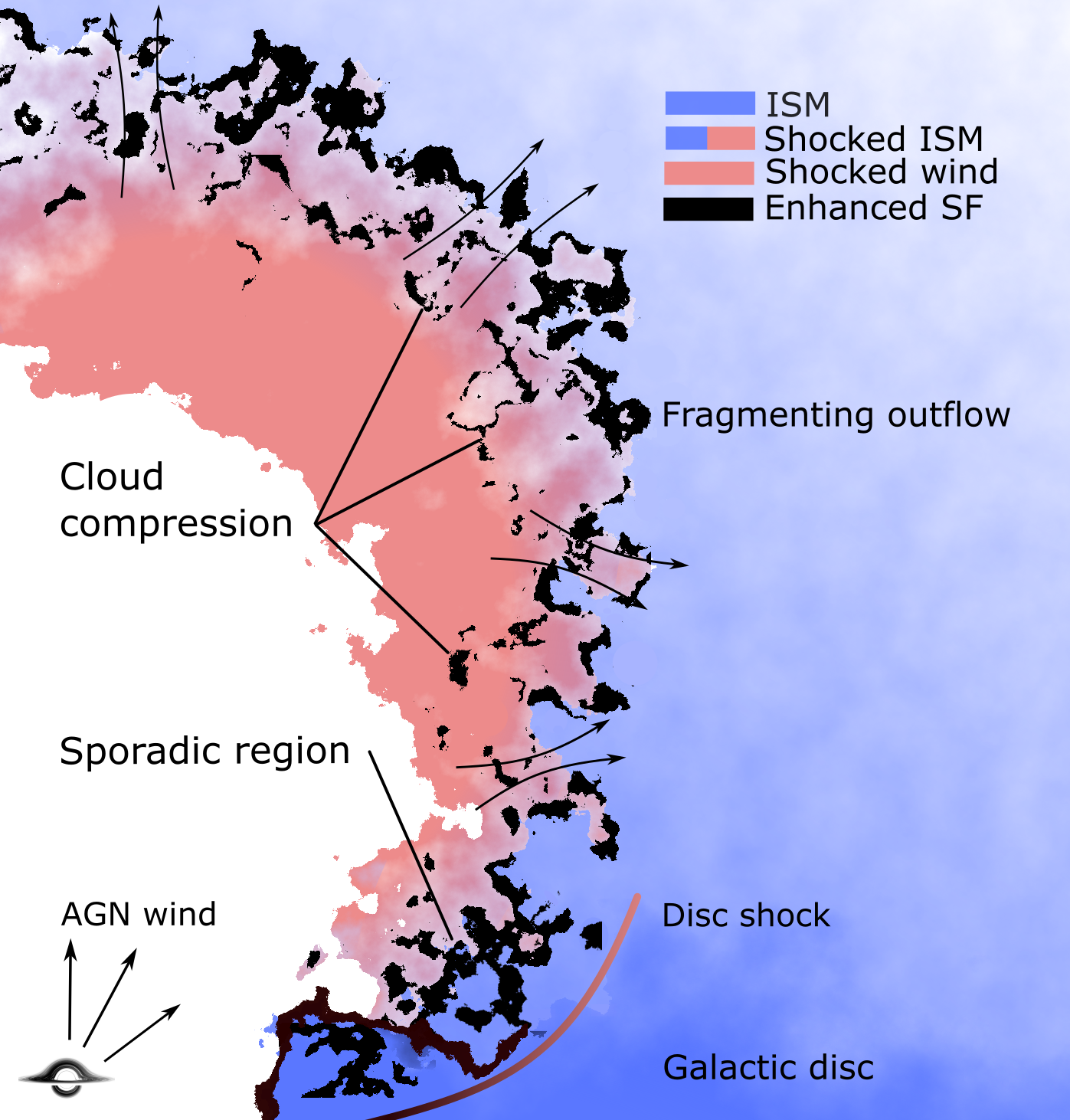}}
        \caption{Diagram of star-formation-enhancing regions. Here we show a schematic of an AGN-driven outflow in a multiphase medium, with the wind filling a cavity surrounded by the shocked wind and shocked ISM. The three possible regions of star formation enhancement are named and indicated by arrows. Outflows can compress stationary clouds directly (for discussion see Sect.~\ref{sec:continuous}). Star formation can be sporadically enhanced in turbulent shear flow between the outflow and the galactic disc (see Sect.~\ref{sec:surroundings}). Finally, outflows can fragment by themselves (see Sect. \ref{sec:Clumpy outflows}).}
        \label{fig:outflow_overview}
\end{figure}

Several mechanisms can transfer the energy liberated during matter accretion on to the SMBH to the surrounding matter. Outflows can form via the action of relativistic winds\footnote{Due to the absence of unified terminology, the terms `wind' and `outflow' are sometimes used interchangeably in the literature. In this work, we use the term `wind' mostly for the quasi-relativistic stream of particles launched by AGN radiation and `outflow' for the mass-loaded slower stream expanding to galactic scales.} \citep{KingPounds:2015:}, radiation pressure on dusty medium  \citep{ArakawaFabian:2022:} or jets \citep{Fabian:2012:}. The relativistic wind, when shocking against the surrounding gas, can provide significant momentum boosts by a factor of ${\sim} 20$ and so is probably the primary contributor to galactic-wide outflows. Wind interaction with the ISM is further divided into two major sub-mechanisms: momentum-driven and energy-driven outflows (for a summary see \citealt{ZubovasKing:2012:a}, Fig. 1). Energy-driven outflows form in diffuse gas, with the main driver being the adiabatic expansion of the shocked wind. In contrast, when the relativistic wind interacts with dense gas, cooling is efficient, leading to momentum-driven outflows. In the case of a multiphase gas, as typically found in galaxies, an energy-driven outflow can form and envelop the embedded cold molecular clouds that are only accelerated by the wind momentum. In the latter case, due to the high density contrast, cloud acceleration is inefficient; the clouds are either compressed or dispersed and create mixing instabilities agitating the flow. As instabilities grow and shocked wind mixes with the shocked ISM, complex multiphase outflows with a wide range of densities, temperatures, and relative velocities between the phases develop. We show a schematic overview of such a multiphase system in Fig.~\ref{fig:outflow_overview}, where we also mark potential regions of star formation enhancement (see Sect.~\ref{sec:cloud-compressing}).

\subsection{Outflow radial properties}
\label{sec:rel_vel}

\begin{figure}
    \centering
    \resizebox{\hsize}{!}{\includegraphics[]{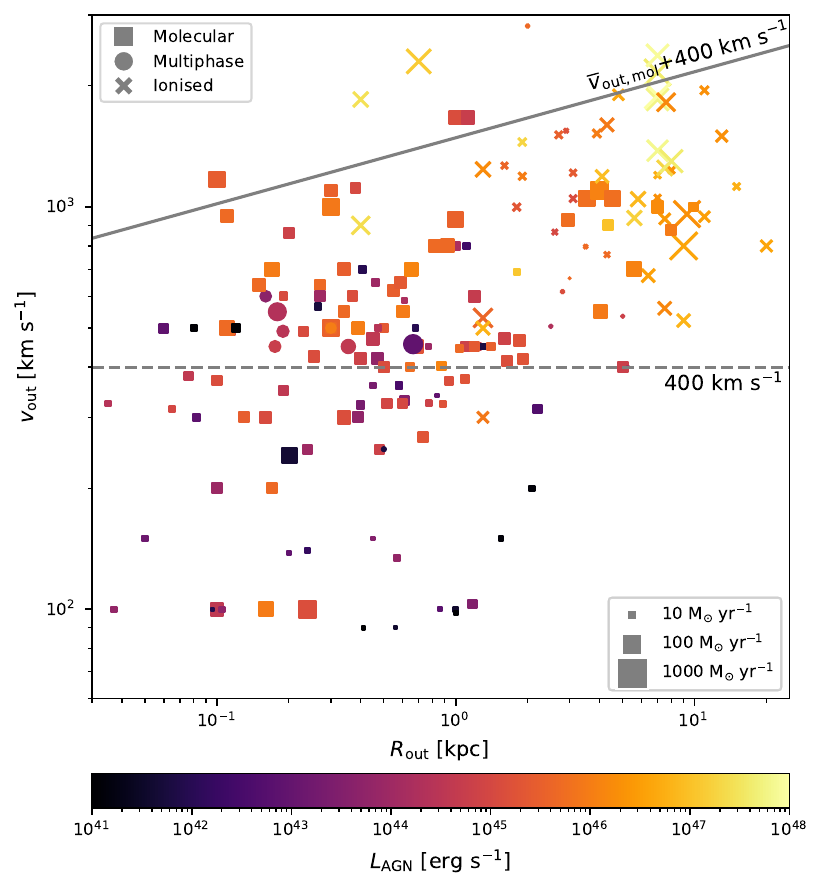}}
    \caption{Compilation of observed outflow velocities and radial distances. The symbol indicates the detected gas phase, symbol size -- mass outflow rate ($\dot M$). The colour indicates the AGN luminosity. The oblique continuous line shows the average $\dot M$-weighted velocity of molecular outflows shifted by $\Delta v_{\rm out} =  400$~km~~s$^{-1}$ . The horizontal dashed line shows the upper velocity limit for outflows simulated in this work. Data aggregated from compilations in \citet{FioreFeruglio:2017:, FluetschMaiolino:2019:, LutzSturm:2020:, ZubovasBialopetravicius:2022:}.} 
     \label{fig:R_out-v_out}
\end{figure}

\begin{figure}
    \centering
    \resizebox{\hsize}{!}{\includegraphics[]{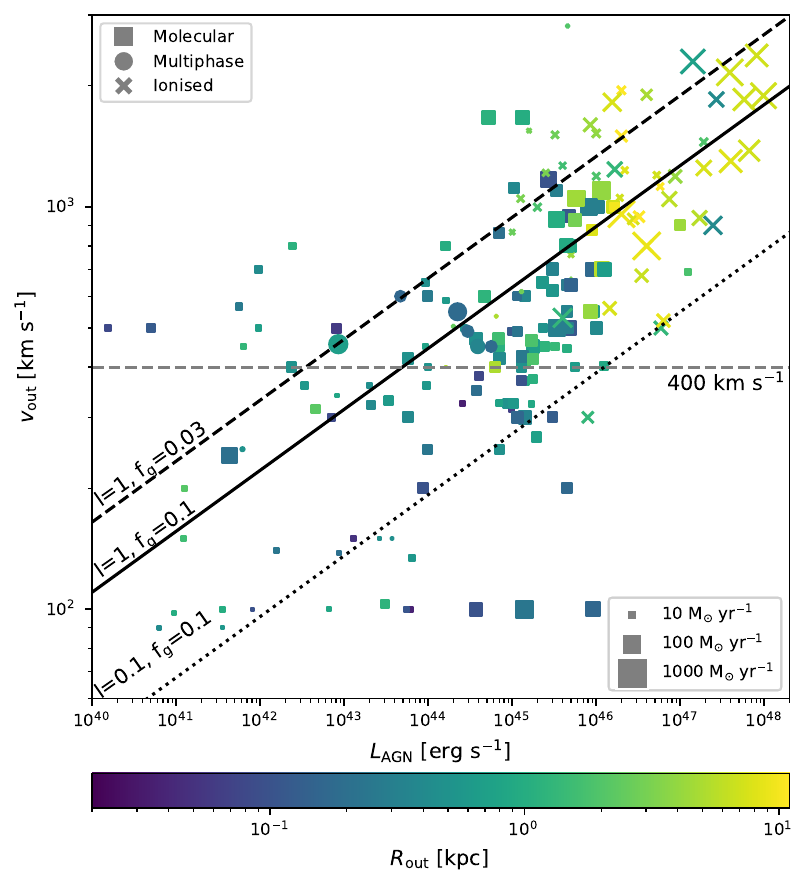}}
    \caption{Same as Fig.~\ref{fig:R_out-v_out} but showing outflow velocity against AGN luminosity. The radius is shown by the colour of the symbols. Oblique lines show the predicted upper outflow velocity limit for SMBH emitting at Eddington ratio $l$ (Eq.~\ref{eq:v_out}) .}
    \label{fig:L_AGN-v_out}
\end{figure}

Galaxy-wide outflows are primarily categorised by the prevalent gas phase, although they are always multiphase. The ratio of the mass contained in different phases and the relative velocity between them depend on AGN luminosity, distance from the nucleus, properties of the surrounding medium and the evolutionary stage of the outflow. We show a compilation (admittedly biased and incomplete) of outflow properties from \citet{FioreFeruglio:2017:, FluetschMaiolino:2019:, LutzSturm:2020:, ZubovasBialopetravicius:2022:} in Figs.~\ref{fig:R_out-v_out} and \ref{fig:L_AGN-v_out}.\footnote{These and all subsequent figures are made using the Matplotlib Python package \citep{Hunter:2007}.} The first figure shows the relation between outflow velocity and radius, with symbol size proportional to the logarithm of the mass outflow rate, symbol type showing the observed outflow phase and colour representing AGN luminosity. In the second figure, we show the dependence of outflow velocity on AGN luminosity, with symbol type and size the same as in the first, while the symbol colour now represents the radius of the outflow. In addition, we plot three lines representing theoretical predictions of outflow velocity under continuous driving, assuming different Eddington ratios ($l \equiv L_{\rm AGN} / L_{\rm Edd}$) and gas fractions ($f_{\rm g} \equiv \rho_{\rm g} / \rho_{\rm total}$) in an isothermal potential. The equation plotted is adapted from \citet[][Eq.~8]{ZubovasKing:2012:b}. We use the observed $M\text{--}\sigma$ relation \citep{FerrareseMerritt:2000:, KormendyHo:2013:}  $M\simeq 3\times10^8 \left(\sigma/200\, {\rm km \,s}^{-1}\right)^{\alpha}$~M$_{\sun}$ with $
\alpha \simeq 4.4$ to eliminate $\sigma$ and the Eddington ratio $l$ to eliminate the SMBH mass, leading to a final form
\begin{equation}
    v_{\rm out} \simeq 1438 \left (\frac{L_{\rm AGN}}{2.3 \times 10^{47}\, {\rm erg\, s}^{-1}}\right )^{2/(3\alpha)} l^{(\alpha-2)/(3\alpha)} \left(\frac{f_{\rm g}}{0.1} \right)^{-1/3} \, {\rm km \, s}^{-1}.
\label{eq:v_out}
\end{equation}
We note that the dependence of this expression on $\alpha$ is very weak.

The plots show several properties of the observed outflow population relevant to our study.  First, there is an overlap between the molecular and ionised outflow velocities and radial distances. The average velocity difference between the phases is of the order of several hundred km~s$^{-1}$, similar to the average velocity of cold outflows; both are also co-spatial within $0.3\text{--}10$~kpc. In this work, we consider outflows with the maximum velocity $v_{\rm out} = 400$~km~s$^{-1}$ (or, equivalently, velocity difference between the hot and cold phases of $\Delta v_{\rm out} = 400$~km~s$^{-1}$). We mark this limit with a horizontal dashed line at $v_{\rm out}$ in Fig. \ref{fig:R_out-v_out} and as a slanted line showing $\overline{v}_{\rm out,mol} + \Delta v_{\rm out}$. Our choice of the maximum upper velocity is arbitrary; however, we infer cloud dispersal and star formation quenching at higher $v_{\rm out}$ or $\Delta v_{\rm out}$. Secondly, slow outflows can be driven even at moderately large AGN luminosities ${\la}3\times10^{46}$~erg~s$^{-1}$. The presence of slow outflows in these AGN hosts can be explained by them having low Eddington ratios (which means they are powered by very massive black holes and reside in very massive galaxies with a strong gravitational potential), the host galaxy being very gas-rich, and/or the AGN driving being intermittent. This last possibility leads to the outflow being driven by an effective luminosity equal to the long-term average AGN luminosity, which may be much lower than the observed instantaneous value \citep{ZubovasNardini:2020:}. The theoretical model we used to draw the lines is based on the assumption of an isothermal density distribution of the gas, which is approximately correct for galactic bulges. In this case, outflow velocity remains constant with radius. In the case of an NFW potential, the velocity first decreases and later increases as the density drops significantly \citep{ZubovasKing:2016:}. This can allow slow outflows to exist in a narrow radial range where the effective velocity dispersion is the greatest. 

Theoretical models provide robust predictions for global structure and radial properties of outflows in smooth medium. However, outflowing material density and temperature on intermediate scales (i.e. tens of parsecs) are less certain as they are determined by the cooling and mixing rates between the gas phases. The sharp discontinuity between shocked wind and shocked ISM might not exist, as it is unstable to KH instabilities \citep{ZubovasKing:2014:}, especially if the medium is inhomogeneous \citep{WardCosta:2024:}. Furthermore, the region close to the discontinuity is heated by thermal conduction, which when combined with mixing causes the layer to swell \citep{WeaverMcCray:1977:}. We estimated the density range of the mixed gas in the layer. For an isothermal sphere, the particle number density is 
\begin{equation}
n_{\rm is} = \frac{\sigma^2 f_{\rm g}}{2 \pi G r^2} \approx 36 f_{\rm g} \ \text{cm}^{-3} \left( \frac{M_{\rm SMBH}}{10^8~\text M_{\sun}} \right)^{2/\alpha} \left( \frac{r}{\text{kpc}} \right)^{-2},
\label{eq:rhon_sis}
\end{equation}
where $r$ is the distance from the AGN. Here we again used the $M\text{--}\sigma$ relation. First we consider galaxies with low bulge gas mass fraction. At a several hundred parsecs from the AGN, outside the dense disc, number density of ISM is several particles per cm$^{3}$. The shocked ISM density is several times above the undisturbed one, of the order of $10$~cm$^{-3}$. The shocked ISM cooling time is longer than the dynamical time, it is Jeans stable and does not fragment. As such, the layer can be considered an extended warm outflow enveloping the molecular clouds. On the contrary, in gas rich galaxies, ambient density is ${>}10$~cm$^{-3}$. The shocked ISM cools and forms a thin, dense layer in which the molecular phase precipitates from the hot outflow \citep{RichingsFaucher-Giguere:2018:, CostaPakmor:2020:}. The shocked and fragmented ISM reduces the covering factor from the position of the AGN, allowing confined high pressure shocked wind to leak and mix with cool gas. In both gas-rich and gas-poor cases, low-density shocked hot wind ($T>10^9$~K, $n \ll 1$~m$_{\rm p}$~cm$^{-3}$) \citep{Faucher-GiguereQuataert:2012:} mixes with cooler ISM regardless of the density of the surrounding medium. The density of the mixed material should be somewhere in between that of the hot wind and the warm envelopes of molecular clouds. In the absence of detailed models, we assumed this density to be ${\sim}1$~m$_{\rm p}$~cm$^{-3}$ and have a wide temperature range of $10^4 \text{--} 10^6$~K. It ablates embedded molecular clumps either compressing or dispersing them. For a more thorough discussion of where and when slow outflows occur (see Sect.~\ref{sec:cloud-compressing}).

\subsection{Molecular clouds}

The vast majority of stars in galaxies form in molecular clouds. In the Milky Way, typical cloud masses range between $10^2\text{--}10^6$~M$_{\sun}$. Smaller clouds are found throughout the periphery of the disc and the bulge, with larger masses towards the disc midplane \citep{Miville-DeschenesMurray:2017:}. The cloud mass function at the high end  is a power law:
\begin{equation}
\frac{dN}{d{\rm ln}M} \propto M^{-\beta},
\label{eq:CloudMF}
\end{equation}
where $N,M$ are the number and mass of molecular clouds, respectively, and the exponent $\beta = 2.0 \pm 0.1$. Despite lower mass clouds being more numerous, the majority of the cold gas in the galaxies is contained within massive $M \geq 10^4$~M$_{\sun}$ molecular clouds.

The properties of many individual clouds are known robustly. Their masses, linear sizes $R$, number densities $n$, and velocity dispersions $\sigma$ tend to follow approximate power-law relations known as Larson laws \citep{Larson:1981:, Miville-DeschenesMurray:2017:, SunLeroy:2018:}:
\begin{equation}
n\propto R^{-p}, \quad M\propto R^q, \quad \sigma\propto R^{r}  
\label{eq.larson},
\end{equation}
with $p\sim 1.0, q\sim 2.1, r\sim0.4$. In our simulations, we assumed that molecular clouds follow these scaling relations exactly (see Sect.~\ref{sec:IC}).

\subsubsection{Cloud compression}

In quiescent galaxies, molecular clouds are in partial equilibrium with the surrounding medium. An outflow surrounding the cloud acts as external pressure and compresses the cold gas. To estimate this effect analytically, we started with a spherical cloud in equilibrium as outlined by \citet{BertoldiMcKee:1992:}, but excluded the magnetic field contribution. We further assumed a constant mean molecular weight $\mu$ and constant specific heat ratio $\gamma$.

The equilibrium condition is defined via the virial equation
\begin{equation}
2(\mathcal{T} - \mathcal{T}_{\rm 0}) + W = 0,
\label{eq:Virial_eq}
\end{equation}
where $\mathcal{T}, W$ are the total kinetic and gravitational potential energies of the system, respectively. External pressure is included via the extra term $\mathcal{T}_{\rm 0} = \frac{3}{2}P_{\rm 0} V_{\rm cl}$. The virial parameter for a spherical cloud is defined as
\begin{equation}
\alpha = \frac{2\mathcal{T}}{|W|} = \frac{5R\sigma^2}{GM} \approx 1.2
\left( \frac{R}{\text{pc}} \right)
\left( \frac{M}{10^3~\text{M}_{\sun}} \right)^{-1}
\left( \frac{\sigma}{\text{km} \ \text{s}^{-1}} \right)^{2}.
\label{eq:Virial parameter}
\end{equation}
Here, $R, M,$ and $ \sigma$ are the radius, mass, and velocity dispersion of the cloud, respectively. With $\alpha \sim 1$ turbulence supports the cloud against gravitational collapse, while clouds with $\alpha < 1$ do not have sufficient kinetic energy and collapse. On the contrary, clouds with $\alpha > 1$  disperse, unless they are bound by external pressure. 

The sources of external pressure can be the surrounding ISM or the outflow. Outflows contribute to external pressure via ram and thermal pressure. Their ratio is
\begin{equation}
\frac{P_{\rm ram}}{P_{\rm th}} =  \frac{K v_{\rm out}^2 \mu \cos^2 \theta}{k_{\rm B} T_{\rm out}}=K \cos^2 \theta \mathcal{M}^2.
\label{eq:pr_ratio}
\end{equation}
Here, $\mathcal{M}$ is the outflow Mach number,  $K(\mathcal M)\la 1$ accounts for the peculiarities of sub, trans, and supersonic regimes \citep{SpreiterSummers:1966:}, and $\theta$ is the angle between the flow direction and the cloud surface normal.

The outflow ram pressure contribution to cloud compression decreases towards the edges of the cloud. In the simplest subsonic case $(\mathcal{M} < 1)$ the outflow impinges on the cloud directly. However, at higher outflow velocities the shock structure becomes more complex. Supersonic interaction $(\mathcal{M} > 1)$ leads to the formation of discontinuities in the outflow, resulting in a bow shock upstream of the cloud \citep{LandauLifshitz:1959:}. In a strong shock ($\mathcal{M} \gg 10$), the evolution becomes practically independent of Mach number \citep{KleinMcKee:1994:} - cloud compression is determined by the gas properties near the cloud surface.

Outflows also drive shocks into the cloud. The shocked gas is compressed, cools rapidly and becomes unstable to fragmentation, resulting in multiple secondary (reflected and refracted) shocks \citep[for shock propagation in clumpy environments see][]{PoludnenkoFrank:2002:}. The resulting cloud and cloudlet compression is non-isotropic and non-homogeneous and depends on the shock strength in the outflow. 

\subsubsection{Cloud evolution timescales}

In the simplest case, without external or internal pressure, a uniform gas sphere collapses under its gravity in a free-fall time that depends only on cloud density $\rho$:
\begin{equation}
t_{\rm ff} = \sqrt{\frac{3 \pi}{32 G \rho}} \approx 1.6 \ \text{Myr} \left( \frac{n}{10^3  \ \text{cm}^{-3}} \right)^{-1/2}.
\label{eq:tff}
\end{equation}
When external pressure is negligible, real clouds evolve on timescales longer than the free-fall time due to internal pressure, which arises due to turbulence, magnetic fields, collapse-induced heating, and (proto-)stellar feedback. As density increases, the free-fall time decreases and locally collapsing regions develop. The typical mass of a collapsing fragment is the Jeans' mass 
\begin{equation}
M_{\rm J} = \frac{\pi c_{\rm s}^3}{6 G^{3/2} \rho^{1/2}} \approx 2 \ \text M_{\sun} \left( \frac{c_{\rm s}}{0.2 \ \rm km \ \rm s^{-1}} \right)^{3} \left( \frac{n}{10^3  \ \text{cm}^{-3}} \right)^{-1/2}.
\label{eq:Jeans}
\end{equation}
Here, $c_{\rm s}$ is the speed of sound. Turbulence-induced overdensities cause further instabilities, so the collapse is not spatially uniform. Cloud collapse and formation of Jeans-unstable gas is further accelerated by external pressure: the external surface of the cloud is compressed first and is susceptible to fragmentation.

The time required for the outflow-induced shock to traverse the cloud is known as the cloud-crushing time:
\begin{equation}
t_{\text{cc}} = \chi^{1/2} \frac{R}{v_{\text{out}}} \approx 9.6 \ \text{Myr} \left( \frac{\chi}{100} \right)^{1/2}
\left( \frac{R}{\text{pc}} \right)
\left ( \frac{v_{\text{out}}}{\text{km} \ \text{s}^{-1}} \right)^{-1},
\label{eq:tcc}
\end{equation}
where $\chi$ is the ratio of cloud density to that of the surrounding medium. As the shock propagates, Kelvin--Helmholtz (KH), Rayleigh--Taylor (RT) and Richtmyer--Meshkov (RM) instabilities form in the interface between the cloud and the ISM \citep{KleinMcKee:1994:, ZhouWilliams:2021:}. The resulting vortices mix the two gas phases, leading to temperature increase and erosion of the cloud's outer layers. The mixing is most prominent where the outflow is tangential to the surface of the cloud. The KH instability growth timescale is approximately 
\begin{equation}
t_{\rm KH} \sim \frac{t_{\rm cc}}{kR}
\label{eq:tKH},
\end{equation}
and the RT instability growth timescale is
\begin{equation}
t_{\rm RT} \sim \frac{t_{\rm cc}}{\left(kR\right)^{1/2}},
\label{eq:tRT}
\end{equation}
where $k$ and $ R$ are the wave number and cloud radius, respectively. In both cases, cloud destruction is dominated by the largest ($kR \ga 1$) wavelengths, which have a characteristic mixing time comparable to the cloud crushing time.

However, this mixing-induced destruction time estimate is not accurate for turbulent clouds undergoing rapid compression. The growth of dense cloudlets is primarily determined by the cooling time 
\begin{equation}
t_{\rm cool} = \frac{2}{3} \frac{k_{\rm B} T}{n\Lambda},
\label{eq:tcool}
\end{equation}
where $\Lambda$ is the volumetric cooling rate. Due to the nonlinearity of the cooling function and outflow pressure anisotropy, we cannot analyse cloud evolution analytically and turn to numerical simulations.

\subsubsection{Star formation}

As the molecular cloud fragments and density of the resulting cloudlets increases, self-gravity becomes the dominant force. The collapse leads to the formation of gravitationally bound dense cores. The characteristic cooling time -- and hence evolution -- of the cores is ${<} 0.1$~Myr. After that, star formation begins \citep{ChevanceKruijssen:2020:}. The star formation efficiency per free-fall time is defined as
\begin{equation}
\epsilon_{\rm ff} = \frac{\dot{M}_*}{M_{\rm cl} + M_* }t_{\rm ff},
\label{eq:epsilojn_eff}
\end{equation}
where $\dot{M}_*$ is the star formation rate and $M_{\rm cl}$ is the cloud gas mass. Typical values of $\epsilon_{\rm ff}$ range between $0.001\text{--}0.1$ depending on spatial scale, with higher values in smaller clouds and/or fragments. When collapsing cores reach density values $10^{6.5}$~cm$^{-3}$, $\epsilon_{\rm ff}$ values increase rapidly, approaching unity \citep{KhullarKrumholz:2019:}. Gas above the density threshold produces stars until eventually stellar feedback and turbulence disperse the cloud (for review, see for example \citet{ChevanceKruijssen:2020:}). The fraction of initial cloud mass converted into stars, known as the integrated star formation efficiency, is $\epsilon_{\rm int} = M_* / (M_{\rm cl} + M_*) \simeq 0.1$ in GMCs in quiescent galaxies \citep{Murray:2011:, KennicuttEvans:2012:}. 

In non-quiescent galaxies, the values of $\epsilon_{\rm ff}, \epsilon_{\rm int}$ have even wider uncertainty.  In addition to self-regulation by stellar feedback, outflows supplement stellar feedback with Mach-number-dependent external pressure (Eq.~\ref{eq:pr_ratio}). Outflows confine, mix and compress the cloud, increasing the variation in $\epsilon$. Due to the complexity of stellar feedback, we limit our investigation to the initial stages of star formation and aim to measure the effects of AGN feedback on the onset of star formation.

\section{Numeric methods}
\label{sec:Numeric methods}

We used {\sc Gadget 4}, a public hybrid SPH/N-body code \citep{SpringelPakmor:2021:} with the pressure-entropy formulation of the equation of motion \citep{Hopkins:2013:} and time-dependent artificial viscosity. The code was chosen due to its scalability and ability to simulate an elongated volume with periodic boundary conditions and gravity. We modelled hydrodynamicsand gravity; radiative processes were simulated via the cooling function. We assumed fully ionised, monoatomic gas and used a specific heat ratio $\gamma=5/3$ with constant mean molecular weight $\mu=0.63$~m$_{\rm p}$. We chose a Wendland C4 kernel \citep{DehnenAly:2012:} with 256 neighbours due to its good performance in analytic tests and modest requirements for computational resources.

\subsection{Initial conditions}
\label{sec:IC}

We simulated a virtual `wind tunnel' -- an elongated box with periodic boundary conditions in all directions. Each simulated system was composed of a stationary cloud and an enveloping outflow. Contrary to the common direct wind or momentum injection into the initially stationary ambient medium,\footnote{See Sect. \ref{sec:wind_shock} for discussion of the differences between the approaches.} we set a uniform initial outflow velocity and artificially heat the ambient gas to match the post-shock conditions.

To investigate the properties of mixed gas, and to enable the analysis of the long-term evolution of the system, we set the length of the box to $l_{\rm box} \simeq v_{\rm out}t_{\rm evo}$, where $t_{\rm evo}$ is the total evolution time.\footnote{We estimated $t_{\rm evo}$ from preliminary low-resolution runs.} The length of the sides perpendicular to the outflow direction was chosen to prevent artificial pressure buildup upstream of the cloud. We set a minimum distance of $30$~pc from the edge of the cloud to the boundary of the box; a complete list of simulated volumes is provided in Appendix~\ref{ap:parameter_table}.

The cloud is a turbulent sphere, with an initially uniform density and temperature of $25$~K. We chose three masses of molecular clouds -- $10^{3}, 10^{4}, 10^{5}$~M$_{\sun}$, and used the Larson relations to determine cloud radii: $R_{\rm cl} = 0.11 \left(M/\text{M}_{\sun} \right)^{0.48}$~pc. Combined with $n_{\rm cl}=1300 \left(R/\rm{pc}\right)^{-1}$~cm$^{-3}$, the initial cloud density values are $n_{\rm cl} = 433, 143, 47$~cm$^{-3}$, respectively. Due to the nature of the mass-radius relationship, each cloud has almost the same column density $\Sigma_{\rm cl} \simeq 40$~M$_{\sun}$~pc$^{-2}$. We set the initial turbulent velocities using $\sigma = 0.27 \left(M/\text{M}_{\sun} \right)^{0.19}$~km~s$^{-1}$, leading to mean values of $1, 1.6, 2.4$~km~s$^{-1}$, respectively. The mass of a single SPH particle was set to $0.1$~M$_{\sun}$, resulting in clouds composed of approximately $10^{4}, 10^{5}, 10^{6}$ particles, respectively. This guaranteed sufficient resolution for mixing with moderate requirements for computational resources.

The rest of the box was filled with a hot, homogeneous outflow with a number density of $1$~cm$^{-3}$; it was simulated with SPH particles of the same mass as the clouds. The outflow has a uniform initial velocity in the $x$ direction (from left to right in figures). We modelled outflows with velocities of $10, 30, 60, 100, 200, 400$~km~s$^{-1}$ and temperatures of $10^4, 10^5, 10^6$~K. These temperatures correspond to sound speeds $c_{\rm s} \sim 15, 47, 148$~km~s$^{-1}$, respectively, so for each outflow temperature, we have both subsonic and supersonic outflows.

Additionally, for each cloud mass, we created a `control' simulation in which the medium surrounding the cloud was stationary and had a temperature of $10^4$~K. Both cloud and outflow parameters were selected to densely fill the parameter space and identify star formation-inducing and quenching regions. In total, we modelled 57 systems (Table~\ref{tab:parameters}). Each system was evolved until fragmentation time (Sect.~\ref{sec:SF}) or until the cloud was dispersed.

\begin{table*}
    \centering
\caption{Summary of the physical parameters.}
\label{tab:parameters}
    \begin{tabular}{cccl}
    \hline\hline
         Parameter&  Unit&  Value& Caption\\
    \hline
         $M_{\rm cl}$&  $\text{M}_{\sun}$&  $10^3, 10^4, 10^5$& Cloud mass\\
         $R_{\rm cl}$&  pc&  $3.0, 9.1, 27.5$& Cloud radius\\
 $n_{\rm cl}$& cm$^{-3}$& $433,143,47$&Cloud number density\\
 $n_{\rm out}$& cm$^{-3}$& 1&Outflow number density\\
 $\chi$& -& $433,143,47$&Density contrast\\
         $T_{\rm cl}$&  K&  $25$& Cloud temperature\\
         $\sigma_{\rm cl}$&  km s$^{-1}$&  $1, 1.6, 2.4$& Cloud velocity dispersion\\
         $v_{\rm out}$&  km s$^{-1}$&  $10-400$& Outflow velocity\\
         $T_{\rm out}$&  K&  $10^4, 10^5, 10^6$& Outflow temperature\\
         $c_{\rm out}$&  km s$^{-1}$&  $15, 47, 148$& Speed of sound of the outflow\\
 $\mathcal M$& -& $0.1-27.1$&Outflow Mach number\\
 \hline
 $T_{\rm norm}$& K& $10^4$&Outflow temperature used for normalisation\\
 $v_{\rm norm}$& km s$^{-1}$& 0&Outflow velocity used for normalisation\\
 \hline
    \end{tabular}
\tablefoot{For a full list, see Appendix~\ref{ap:parameter_table}.}

\end{table*}

\subsection{Turbulence}
\label{sec:Turbulence}
To generate a realisation of turbulent velocities, we started with a uniformly spaced lattice, totalling 256$^3$ points in Fourier ($k$) space. At each point, the amplitudes of the generating field are determined by the power law
\begin{equation}
    P  \equiv  \langle |v_k|^2 \rangle \propto k^{-11/3}.
        \label{eq:pspec1}
\end{equation}
For each grid point, we sampled a random complex number from a bi-variate Gaussian distribution as outlined in \citet{DubinskiNarayan:1995:}. The resulting field amplitudes at a point are Rayleigh distributed with uniform phase distribution from 0 to $2\pi$.

The resulting velocity field, formed by such a process, is purely compressive. However, various compositions of the turbulent field can be recovered using a projection operator \citep{FederrathKlessen:2008:}:

\begin{equation}
    V_{ij}^{\tau}(k) = \tau V_{ij}^{\perp} + (1-\tau)V_{ij}^{\parallel} =\tau \delta_{ij} + \left(1 - 2\tau \right) \frac{k_i k_j}{|k|^2}.
        \label{eq:projection_op}
\end{equation}

Here, $V_{ij}^{\tau}$ is a projection tensor in the Fourier domain, and $\tau \in [0\text{--}1]$ is the ratio of solenoidal turbulent energy to total. In the second part of  Eq.~\ref{eq:projection_op}, we decomposed the projection tensor into a sum of transversal $V_{ij}^{\perp}$ and longitudinal $V_{ij}^{\parallel}$ operators and finally rewriten them as the sum for vector components in Fourier space, where $\delta_{ij}$ is the Kronecker delta as usual. We tested the effect of $\tau$ on fragmentation and found no significant differences for an expected range $0.6\leq \tau \leq1.0$ in realistic clouds \citep{GinsburgFederrath:2013:}. As a result, we selected purely solenoidal turbulence ($\tau = 1$) for this work.

Finally, the velocity field components ($v_x, v_y, v_z$) were calculated by an inverse Fourier transform. The turbulent velocity is the real part of the transformation. Velocity components of individual SPH particles were calculated by linear interpolation.

\subsection{Cooling function}
\label{sec:Cooling}

We estimated radiative cooling rates with a cooling function as presented in \citet{KakiuchiSuzuki:2024:}. For low temperatures (i.e. $T < 10^4$~K) it uses the results by \citet{KoyamaInutsuka:2000:, KoyamaInutsuka:2002:}, whereas at higher temperatures, it uses a fit to \citet{SutherlandDopita:1993:}. (For the explicit form of the cooling function, see Appendix~\ref{ap:cooling}.)

\begin{figure}
     \centering
        \resizebox{\hsize}{!}{\includegraphics[]{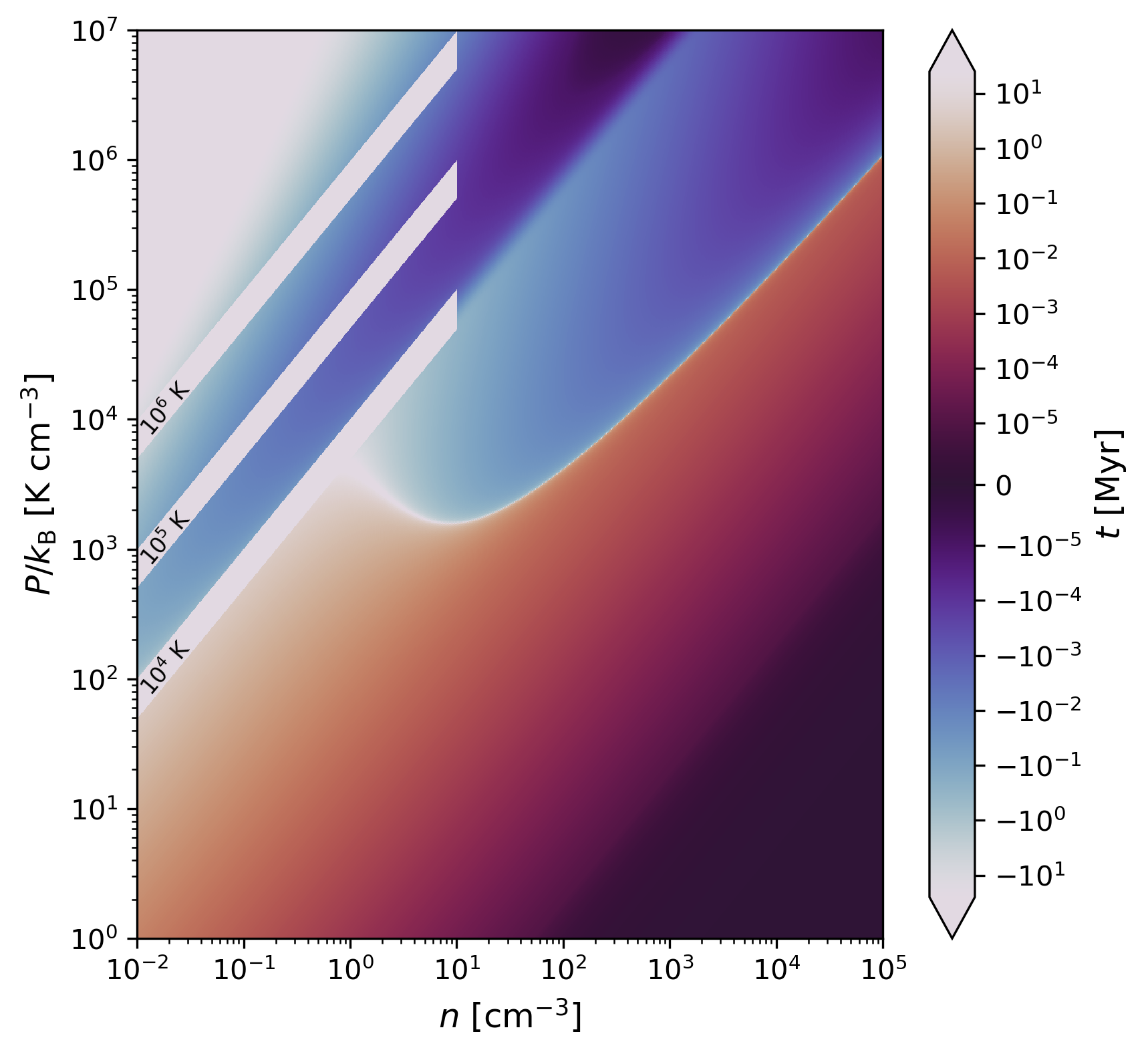}}
        \caption{Heating and cooling (negative) times. Artificially heated outflow regions are shown as oblique grey stripes, with outflow temperatures indicated on the plot. }
        \label{fig:cool-function}
\end{figure}

We neglected thermal conduction, assuming it is either small compared to radiative cooling, or the flux is saturated, or evaporation times in the classical regime are long compared to the evolution times of our systems \citep{Spitzer:1978:}.

Real outflows are expected to maintain approximately constant temperature, at least during the AGN phase. To approximate this, we artificially heated the ambient gas: for gas particles with number densities $n < n_{\rm thresh} =10$~cm$^{-3}$ and temperatures in the range of $T_{\rm out}/2 < T < T_{\rm out}$, the value of $\Lambda$  (Eq.~\ref{eq:Cooling_model}) was set to zero. We tested several sets of density and temperature thresholds with little impact on the results (see also Appendix~\ref{ap:Tn_diag}). The modified cooling function ensures a warm-to-hot outflow with constant temperature and pressure and has little effect on the total mass of warm gas produced by mixing.

We show the heating and cooling time dependence on gas density and pressure in Fig.~\ref{fig:cool-function}. The modified cooling regions are highlighted in grey, with corresponding outflow temperature labels. The thin grey curve starting around $n \sim 1$~cm$^{-3}$ and $P/k_{\rm B} \sim 5\times10^3$~K~cm$^{-3}$ is the equilibrium state of the gas.

\subsection{Star formation}\label{sec:SF}

We adopted a simple prescription to track star formation: we assumed Jeans unstable gas with number densities $n > 10^{6}$~~cm$^{-3}$ rapidly collapses into protostellar fragments. As soon as an SPH particle satisfied both conditions, it was instantaneously converted to a star particle and subsequently interacted with the rest of the system only through gravity. We evolved each system until the total mass of star particles reached $f_* = 0.04$ of the initial cloud mass. We labelled this moment as the `fragmentation time', $t_{\rm frag}$.

To account for different evolution timescales of clouds with different masses, we defined a normalised dimensionless fragmentation time as
\begin{equation}
    t_{\rm norm}= \frac{t_{\rm frag}}{t_{\rm frag(ref)}}
    = \frac{t[M_{\rm cl}, \ v_{\rm out}, \ T_{\rm out}]}
    {t[M_{\rm cl}, \ 0 \ {\rm km} \ {\rm s}^{-1}, \ 10^4 \ {\rm K}]},
    \label{eq:t_norm}
\end{equation}
where $t_{\rm frag}$ is the fragmentation time of a given system, while $t_{\rm frag(ref)}$ is the fragmentation time of the control simulation -- a cloud of the same mass in stationary $10^4$~K ambient medium.

\section{Results}
\label{sec:Results}

We first present the evolution of gas morphology, followed by the evolution of the mass of gas of different phases and finally move on to the onset of fragmentation and initial fragmentation rates. We use a Cartesian coordinate system with outflow velocity in the positive $x$ direction in all density maps.

\begin{figure*}
     \centering
        \includegraphics[width=1\textwidth]{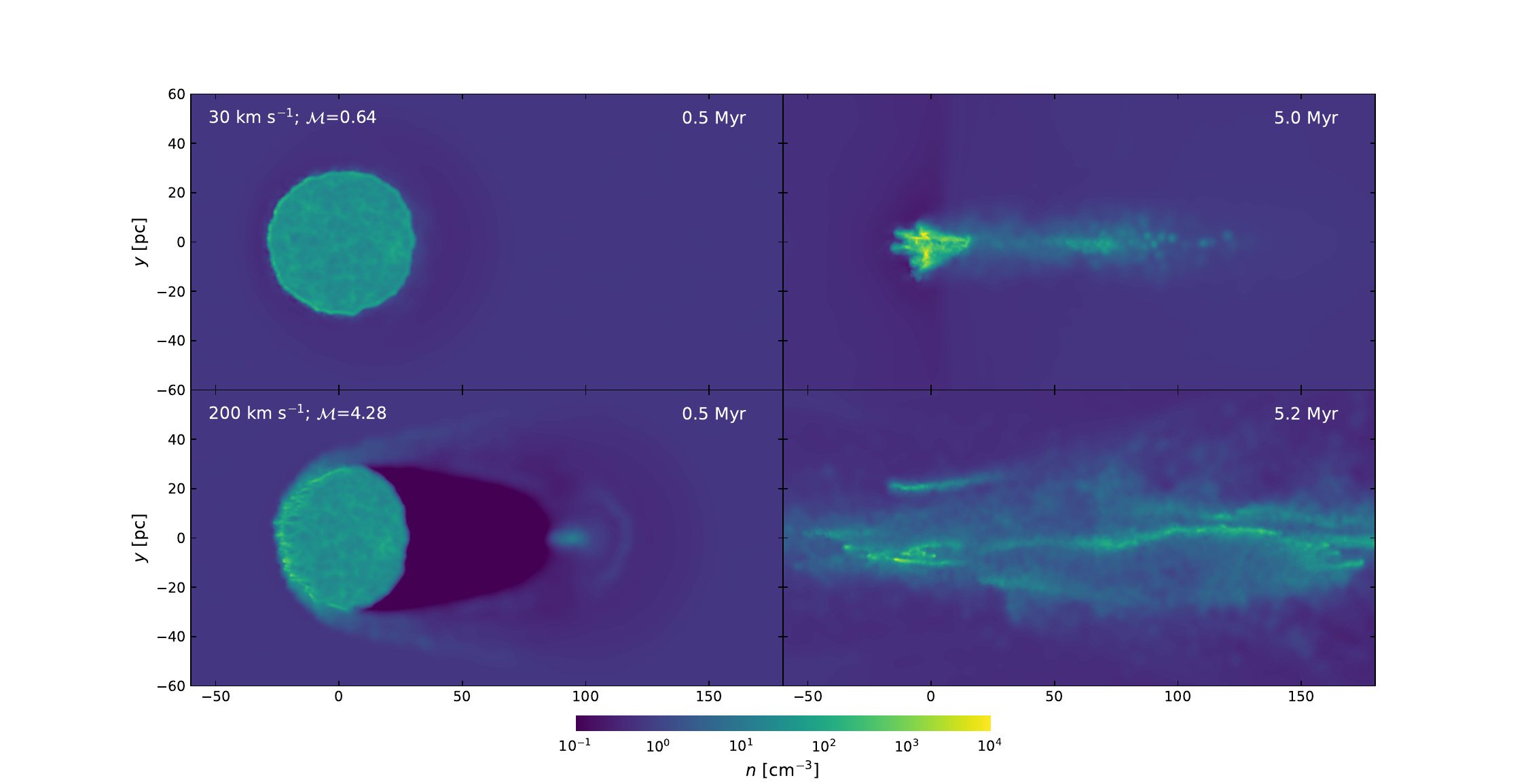}
        \caption{Particle density slices of a $10^5$~M$_{\sun}$ molecular cloud embedded in a $10^5$~K outflow. The left column shows the system's state after 0.5~Myr, the right one -- at $t_{\rm frag}$. Outflow velocity (Mach number), indicated at the top left of each panel is 30~km~s$^{-1}\, (\mathcal{M}=0.64)$ in the top row and 200~km~s$^{-1}\, (\mathcal{M}=4.28)$ in the bottom one. Coordinates are centred on the cloud centre of mass, with outflow velocity in the positive $x$ direction. For a complete set of clouds in $10^5$~K outflows and equivalent plots of clouds in $10^4$~K and $10^6$~K outflows, see Appendix~\ref{ap:density_maps}.}
        \label{fig:EVO}
\end{figure*}

\subsection{Morphology}
\label{sec:Morphology}

The outflow and cloud interaction can be split into four phases. First, as the outflow impacts the cloud, an inner cloud shock forms. During the second stage, the cloud is compressed, mostly in the direction of the outflow. As the shock traverses the cloud, fragmentation begins. The final stage is cloud destruction due to the mixing of cloud and outflow material, which allows instabilities to develop and produces eddies. The interaction of subsonic and supersonic outflows with the cloud differ in their details - we present them below.

\subsubsection{Subsonic outflow}
In subsonic interaction (i.e. simulations with $v_{\rm out} < c_{\rm out}$; top two rows of Fig.~\ref{fig:EVO}), the flow over the surface of the cloud is smooth and follows the contours of the cloud. The outflow ram pressure is low, so pressure anisotropy is weak. The cloud is more strongly deformed by the Venturi effect, that is, lower pressure caused by accelerated flow along the edges of the cloud. However, this does not result in major morphological changes and cloud compression can be considered isotropic. A dense collapsing shell forms, which sweeps the turbulent cold cloud gas, leading to accelerated growth of overdensities in the shell.

The dense collapsing shell also inhibits large-scale instability-induced mixing. On smaller scales, the large density contrast and efficient cooling reduces shear flow mixing even further. Therefore, the cloud is confined and retains the majority of its initial mass. The cloud's interior is isolated from the outflow and evolves independently.\footnote{We note that this aspect of cloud evolution is markedly different from that in the `blob test' \citep{AgertzMoore:2007:}, which our setup superficially resembles. It arises because of cooling, which is neglected in the adiabatic tests.} Turbulence causes prominences to extend from the cloud, which can be blown away by the outflow. However, their removal is also inefficient due to the high-density contrast and so the resulting mass loss is negligible. At later stages, the shell fragments due to self-gravity and is no longer able to shield the cloud interior from the flow. As the outflow pierces the cloud shell, the mass of the turbulence-induced clumps (${>} 50$~M$_{\sun}$) determines whether they are dispersed. Eventually, the clumps cool and contract due to self-gravity, their density reaches the star formation threshold and star particles begin to form (Sect.~\ref{sec:SF}).

\subsubsection{Supersonic outflow}
In the supersonic models (third and lower rows of Fig.~\ref{fig:EVO}), a bow shock upstream of the cloud is essentially indistinguishable from the surface of the cloud due to the fast cooling of shocked material. In hot outflows $T = 10^6$~K, cooling is less efficient, thus a prominent shock forms upstream of the cloud  (see Appendix \ref{ap:density_maps}). The reverse shock, compressing the opposite side of the cloud is observed only in the low supersonic regime. The cloud shock produces fragmented multi-phase gas with  $t_{\rm ff} \simeq t_{\rm cool} < t_{\rm KH} < t_{\rm cc}$ in the densest clumps. The clumps obstruct the flow, however, the ram pressure is high enough to pierce low-density regions resulting in the formation of dense pillars in the direction of the outflow. 

Fragmentation of the shell and shocked cloud material produces multiple weaker shocks and mixing instabilities inside the cloud, with a typical size of a few parsecs. Simultaneously, as the shock traverses the cloud, the dense pillars and diffuse channels penetrate within until they traverse the cloud completely. When this happens, fast outflow material starts streaming through the cloud. Ablation by the outflow results in an increased area of contact between the outflow and cloud material, further increasing their mixing. The relative timing of fragmentation and channel formation determines the ultimate fate of the system. If the channels penetrate the cloud before the formation of dense self-gravitating cloudlets, the cloud is dispersed and star formation quenched. Conversely, if self-gravitating cloudlets form earlier, the outflow material filling the channels further compresses the cloudlets and star formation is enhanced.

At the edges of the cloud, KH and RT instabilities form large eddies with a circulation diameter comparable to the radius of the cloud. These eddies produce mixing regions behind the cloud, separated by a low-density cavity in between (bottom three rows of Fig.~\ref{fig:EVO}, left column; see also Sect.~\ref{sec:mass_cool}). These large-scale mixing processes are suppressed in later stages by outflow material piercing the cloud, filling the cavity behind the cloud, and inhibiting further growth of mixing instabilities (right panels of Fig.~\ref{fig:EVO}).

\subsection{Mass and dynamics of cool gas}
\label{sec:mass_cool}

The competing processes of shock-induced heating and radiative cooling of the dense material lead to constant changes in the material comprising the cloud. In addition, ram pressure and self-gravity compete to move the cloud material in different directions, facilitating or impeding cloud destruction. The net effect of all these processes can be seen by considering the evolution of the cold gas mass.

We define cold gas, as that with a number density ${>} 10$~cm$^{-3}$ and temperature ${<} 10^3$~K, while the rest is considered to be part of the outflow. Both thresholds are chosen to be consistent with the threshold values in the cooling function (Sect.~\ref{sec:Cooling}). The precise threshold values have little impact on our results, since in most cases the cloud and outflow phases are well separated both in density and in temperature (see Appendix~\ref{ap:Tn_diag}).

Figure~\ref{fig:all_mass} shows the evolution of cloud mass (solid lines), and the difference between converging and diverging cold gas mass (dashed lines) of $M_{\rm cl} = 10^5$~M$_{\sun}$ simulations. Both quantities are given as ratios with the initial cloud mass. Cloud mass decreases due to compression heating and evaporation. Conversely, efficient radiative cooling results in the growth of cold gas mass. Cold mass also grows due to thermal instability in the swept-up hot outflow medium, provided the gas can cool efficiently. Measuring the difference between converging and diverging cold gas mass helps elucidate which process dominates. Positive values of the difference mean that the cloud itself is growing (the net flux of cool gas is towards the cloud's centre of mass), while negative values indicate cloud dispersal.

\begin{figure*}
     \centering
        \includegraphics[width=0.9\textwidth]{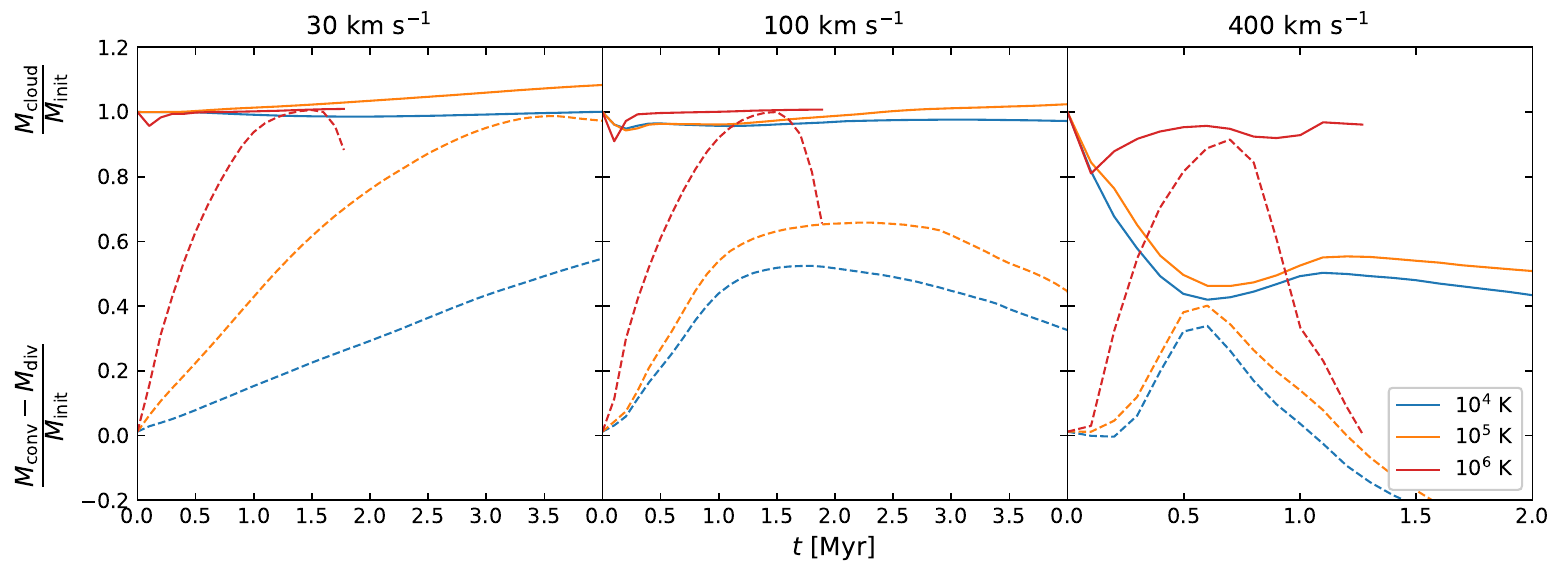}
        \caption{Evolution of cold gas mass. We show the ratio of cold gas mass to initial cloud mass (solid line) and the ratio of the difference between converging and diverging cold gas mass to the initial cloud mass (dashed line) for the $M_{\rm init} = 10^5$~M$_{\sun}$ simulations. The outflow velocity increases from left to right; colours show the outflow temperature as indicated in the legend. (For equivalent plots of clouds with lower masses, see Appendix~\ref{ap:cold_gas}.)}
        \label{fig:all_mass}
\end{figure*}

The mass of the cold gas remains approximately constant for velocities ${\leq} 100$~km~s$^{-1}$ independently of outflow temperature. Faster outflows, produce two stages in mass evolution. Initially, the mass decreases as cloud material is heated above the cold gas threshold due to compression by the propagating shock wave. The shocked gas fragments and is accelerated by the outflow, increasing the mass of the in-falling gas. After the cloud shock traverses the cloud's centre of mass (a peak in the converging/diverging mass graphs in Fig.~\ref{fig:all_mass}), the second stage begins. During this stage, the fragmented gas mixes with the outflow, and the in-falling gas fraction decreases. Some of the mixed gas cools down, adding to the cold gas mass. As the shock fully traverses the cloud (an increase in the mass ratio starting ${\sim} 1$~Myr at $400$~km~s$^{-1}$), the cold gas mass begins to decrease again as the cloud starts to disperse. This effect is absent in the $10^6$~K outflows, where the cold gas mass remains fairly constant due to the high thermal pressure confining the cloud, thus preventing dispersal.

As the shock propagates through the cloud and the gas fragments, outflow material streams through the clouds, enhancing mixing and dispersing the cloud. This is indicated by negative values of the converging-diverging gas mass difference. The destruction of the cloud produces elongated diffuse warm clouds that retain a significant fraction of the initial molecular cloud mass. 

At supersonic velocities, a cavity with a very low number density ($n \ll 1$~cm$^{-3}$) forms behind the cloud. Outflows with temperatures ${\ga} 10^5$~K eventually crush the cavity behind the cloud leading to thermal instability that causes cool gas to `precipitate' \citep{InoueInutsuka:2008:}. This effect is most prominent at $100\text{--}400$~km~s$^{-1}$ outflow velocities and $10^5, 10^6$~K outflow temperatures. The velocity of the cloudlets is similar to the velocity of the outflow, and therefore it increases the mass of the diverging gas and simultaneously contributes to the mass growth of the cold gas. The gas in the cloudlets does not fragment (see Sect.~\ref{section:Fragmentation}) but initiates further thermal instabilities in the outflow.

\subsection{Fragmentation}
\label{section:Fragmentation}

As explained in Sect.~\ref{sec:SF}, we use a normalised fragmentation time (Eq.~\ref{eq:t_norm}) as the primary metric for star formation quenching or enhancement. The fragmentation times in the three `control' simulations with stationary, warm ($10^4$~K) ambient medium are $2.43, 5.50$ and $7.18$ Myr for $10^3, 10^4$ and $10^5$~M$_{\sun}$ clouds, respectively. 

\begin{figure}
    \centering
    \resizebox{\hsize}{!}{\includegraphics[]{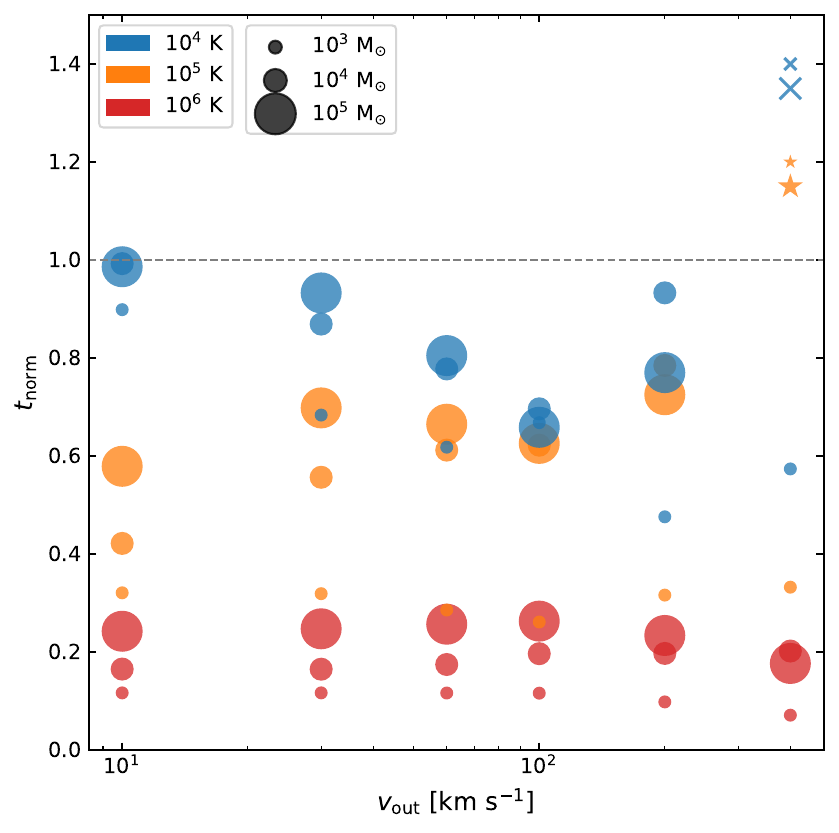}}
    \caption{Normalised fragmentation times. Symbol size shows cloud mass, and colours show the outflow temperature, as given in the legend. Circles show systems that reached the fragmentation time, and crosses indicate systems where the cloud is dispersed before forming any star particles. Star symbols represent systems where a few star particles form but the cloud is dispersed before reaching the adopted fragmentation criterion. The vertical position of the symbols for non-fragmented systems is arbitrary. The dashed horizontal line separates star formation enhancing ($t_{\rm norm} < 1$) and quenching regions.}
    \label{fig:modeliu_tinklelis}
\end{figure}

Fragmentation occurs either in the shocked cloud gas or in the undisturbed cloud within dense turbulence-induced filament knots. Shocked gas is heated to several thousand K and fragments as it cools, and unshocked cloud gas is compressed by turbulent velocities before the arrival of shock. Lower shock velocity and larger cloud radius prolong the initial turbulence-induced filament formation phase resulting in denser clumps.\footnote{We use clustering algorithm DBSCAN \citep{10.1145/3068335} to determine the properties of the clumps at fragmentation time.} In $10^4, 10^5$~K outflows, the mean mass of the resulting cloudlets is ${\sim} 300$~M$_{\sun}$ with a mean radius of ${\sim} 1.0$~pc. Clouds of initial mass $10^5$~M$_{\sun}$ produce ${\sim} 100$ cloudlets with a few reaching $10^4$~M$_{\sun}$, fragmentation occurs primarily in the most massive ones. The rest of the gas in the cloud is too diffuse and uniform to be considered a cloudlet. Fragmentation of the $10^4$~M$_{\sun}$ clouds is similar, except for the lower overall count (${\sim} 30$) and lower mass ($10^3$~M$_{\sun}$) of the largest cloudlet. The lowest-mass clouds hardly fragment at all, usually collapsing into a single or several cloudlets with a cumulative mass of $500\text{--}800$~M$_{\sun}$. Despite the differences in the initial density, the fragmentation of the clouds is self-similar, and the minimum size of the cloudlets formed is determined by the cooling rates. Contrary to the cooler ones, $10^6$~K outflows rapidly compress the cloud, and the turbulence-induced overdensities are aggregated into several massive  ($>10^4$~M$_{\sun}$) clumps. At supersonic velocities, the clouds shatter - the clumps completely separate from each other.

Figure~\ref{fig:modeliu_tinklelis} shows the normalised fragmentation times for all simulations. In the vast majority of our simulations, fragmentation is enhanced, that is, the compressive effect of outflow temperature is higher than the disruptive effect of its kinematics. As expected, higher outflow temperature results in stronger enhancement. In extreme cases, for the smallest clouds embedded in hot ($10^6$~K) outflows, the fragmentation time is up to a factor ${\sim} 10$ shorter than in the control runs, with more massive cloud fragmentation enhanced by a factor ${\sim} 5$. Even at the highest velocities we have simulated, the thermal pressure of hot outflow crushes the cloud, preventing dispersal. (For a full list of numerical values, see Appendix~\ref{ap:parameter_table}.)

In $10^5$~K outflow simulations, cloud compression is less effective, roughly by a factor of $2\text{ to }3$. Again, clouds of lower mass are more strongly affected and fragment independently of the kinematic pressure. At ${\geq} 100$~km~s$^{-1}$,  more massive clouds are broken up into distinct components with masses similar to the clouds of the lowest mass, which then evolve roughly independently of one another. At the highest outflow velocity, massive clouds are dispersed before reaching the fragmentation threshold. The lowest mass clouds survive and fragment at all velocities. All clouds form some fragments before being dispersed (indicated by star symbols in Fig.~\ref{fig:modeliu_tinklelis}).

In warm outflows ($10^4$~K), the outflow thermal pressure is similar to that of the undisturbed ISM - in other words, the outflow is close to pressure equilibrium with the cloud. This allowed us to effectively isolate the influence of ram pressure even at velocities as low as $30$~km~s$^{-1}$. In such low-velocity simulations, fragmentation is enhanced by up to a factor of $1.5$ in massive clouds. The enhancement is even stronger at $v_{\rm out} = 100$~km~s$^{-1}$. At $200$~km~s$^{-1}$, the normalised fragmentation time of the massive clouds approaches unity, indicating a transition to a cloud dispersal regime. At $400$~km~s$^{-1}$, all but the lowest-mass clouds are completely dispersed and do not fragment.

In the majority of simulations with $v \leq 100$~km~s$^{-1}$, fragmentation is rapid, lasting ${\sim} 0.1$~Myr from the moment the first fragments appear until fragmentation time is reached. However, at higher outflow velocities (${>} 100$~km~s$^{-1}$), there is competition between cloudlet destruction and fragmentation, leading to prolonged fragmentation spanning ${>} 1$~Myr at the highest velocities.

Overall, our results show that there is a region of outflow and cloud parameter space where outflow passage enhances fragmentation (and, most likely, subsequent star formation). This parameter space is defined by threshold velocities of ${>} 400$~km~s$^{-1}$ at an outflow temperature of $10^6$~K, and ${<} 400$~km~s$^{-1}$ at $10^4, 10^5$~K.

\section{Discussion}
\label{sec:Discussion}

We simulated the interaction of an isolated turbulent molecular cloud and a warm outflow, and identified a region in the parameter space of cloud mass, outflow velocity, and temperature in which gas fragmentation, and presumably subsequent star formation, is enhanced. Slow outflows with velocities ${\leq} 200$~km~s$^{-1}$ induce or enhance star formation in a wide temperature range, increasing gas consumption rates. Faster outflows disperse the massive clouds and quench star formation unless the outflow temperature reaches $10^6$~K, in which case the velocity threshold for destruction rises to ${>} 400$~km~s$^{-1}$.

We observed, that star formation enhancement is likely in regions where outflows are moving at low-to-moderate velocities relative to the dense gas clouds. In the next section, we provide an overview of such scenarios. We then discuss contributions from other star-formation-enhancing or quenching processes that can mask the effects of positive AGN feedback. Later, in Sect.~\ref{sec:Star formation}, we briefly overview the expected kinematics of stellar populations formed in this regime, then compare our results with several other works in Sect.~\ref{sec:comparison} and lastly address the shortcomings of our models in Sect.~\ref{sec:caveats}.

\subsection{Fragmentation-enhancing scenarios}
\label{sec:cloud-compressing}

\subsubsection{Compression of stationary clouds in continuous outflows}
\label{sec:continuous}

We first consider the compression of clouds with initially negligible radial velocity by the passage of an AGN-driven outflow. The simplest viable model of AGN outflow propagation is a spherically symmetric energy-driven wind model \citep[e.g.][]{KingPounds:2015:}. Assuming constant AGN luminosity and an isothermal density profile, the outflow velocity quickly reaches a constant value and maintains it at least until it clears the bulge. The velocity also determines whether this duration is long enough to compress the cloud and significantly enhance its fragmentation rate. In the simplest, fully adiabatic, case, the thickness $l_{\rm out}$ of the shocked ISM layer is ${\sim} 1/4$ of the total outflow radius \citep{ZubovasKing:2014:}. So the cloud is compressed for a duration $t_{\rm compr} \sim l_{\rm out} / v_{\rm out} \sim R_{\rm out} / \left(4 v_{\rm out} \right)$. Compression is significant only when it lasts for at least ${\sim} t_{\rm frag}$ and preferably longer. In our simulations, $t_{\rm frag}$ ranges from ${\sim} 0.3$~Myr for the smallest highly compressed clouds to ${>} 3$~Myr for the slightly compressed most massive clouds. Considering a range of $v_{\rm out}$ between $10\text{--}400$~km~s$^{-1}$, we get a minimum thickness required for sustained compression to be $0.005$~kpc $< l_{\rm out} < 1.2$~kpc. Equivalently, the required outflow radii have a range of $0.02$~kpc $< R_{\rm out} < 4.8$~kpc.

Observational evidence (see Fig.~\ref{fig:R_out-v_out}) suggests, that the upper half of this range may occur only rarely, if ever. The maximum radii of outflows with $v_{\rm out} < 400$~km~s$^{-1}$ are ${\sim} 2.5$~kpc, and the majority of such outflows are found within the central kiloparsec. In all cases, such outflows only exist in AGNs with luminosity $L_{\rm AGN} < 10^{46}$~erg~s$^{-1}$ (Fig.~\ref{fig:L_AGN-v_out}). This suggests that star formation enhancement should also be expected mainly in the central parts of galaxies with low-power AGNs, rather than further out - in the halo.

If we consider a more realistic setup, the picture described above becomes more complicated (see Fig.~\ref{fig:outflow_overview}). Different gas density profiles and AGN luminosity variability on timescales comparable to outflow expansion induce changes in outflow velocity and/or temperature. Additionally, even if the outflow expands at a constant velocity, its density, and hence pressure, decreases with time \citep{ZubovasKing:2016:}. We can expect that a decreasing pressure would have a lower effect on the cloud than a constant one, but we refrain from further speculation since we did not simulate such a scenario. We intend to investigate how a varying outflow affects the star formation rate in future work.

One more complication is the cooling of the shocked ISM. If it is rapid, the predicted outflow thickness decreases, potentially by a large factor \citep{RichingsFaucher-Giguere:2018:}. In that case, a typical stationary cloud first traverses the ISM shock and, soon after, the contact discontinuity. The passage is less disruptive since the pressure is the same on both sides. However, after traversal, the cloud is embedded in the shocked AGN wind, with extremely high temperatures reaching ${>} 10^9$~K. As the shocked fragmented cloud is engulfed in the hot wind, thermal conduction can no longer be ignored, leading to evaporation of cloudlets in ${\ll} 1$~Myr \citep{CowieMcKee:1977:}. From this, we conclude that rapidly cooling outflows are unlikely to enhance star formation in dense pre-existing ISM clouds. However, they can lead to star formation by fragmenting themselves (see Sect.~\ref{sec:Clumpy outflows} below).

\subsubsection{Sporadic fragmentation in clumpy galactic discs}
\label{sec:surroundings}

We now discuss viable regions in or near the galactic disc where the velocity difference between hot and cold phases falls within the star-formation-enhancing region of the parameter space considered in this work. This mode depends on many circumstances of ISM distribution and instabilities along the sides of the outflow, so we call it `sporadic' (see Fig.~\ref{fig:outflow_overview}).

The vast majority of outflows in active galaxies are bi-cones \citep{NevinComerford:2018:}.  The dense disc gas collimates the outflow and a layer with density and velocity gradients between the conical outflow and its surroundings forms. The shear flow in these layers creates hydrodynamical instabilities, increases turbulence and thus can compress the clouds. This happens essentially independently of the geometry of the outflow as it is being launched or independently of the driving mechanism.  Due to the collimation, as well as the increasing mass of swept-up and mixed gas, the outflows have diminishing effects on the galactic disc with increasing radial distance. This has been confirmed by the results from the MaNGA survey \citep{IlhaRiffel:2019:}: they show that low luminosity AGNs alter gas kinematics only within $1\text{ to }2$~kpc and does not affect the gas outside this region. The intrinsic velocity of the gas in the cones is ${\sim} 300$~km~s$^{-1}$ and the cones extend to $3.4 \pm 1.8$~kpc. The observed velocities fall within the star-formation-enhancing region of our simulations. Therefore, star formation is likely enhanced along the cone boundary.

Such a sporadic star formation may be the initial link in the positive feedback chain as outflows can also enhance star formation by compressing the galactic disc as proposed by  \citet{ZubovasNayakshin:2013:b}.  Such action extends the range of positive feedback from shear flow layers of the outflow to the galactic plane, possibly increasing star formation rates by up to an order of magnitude \citep{ BieriDubois:2016:}. While such proposed star-formation-enhancing modes are expected from theoretical and numeric models, the observational evidence remains unclear. This discrepancy can be explained by the high-Eddington ratio AGN phase being short compared to the outflow dynamic time. We address the implication of different timescales in Sect. \ref{sec:Fossil outflows}.

\subsubsection{Fragmentation of clumpy outflows}
\label{sec:Clumpy outflows}

Outside the galactic disc, outflows sweep and compress tenuous ISM and form shells of outflowing material. In gas-rich ($f_{\rm g} \ga 0.1$) systems, some fragmentation of the shell appears inevitable \citep{NayakshinZubovas:2012:, Scannapieco:2017:, RichingsFaucher-Giguere:2018:}; this is accompanied by rapid cooling of the fragments \citep{ZubovasKing:2014:}. Observations support the fragmenting outflow scenario as most AGN-driven outflows are dominated by molecular gas at a lower bolometric luminosity $L_{\rm AGN} <10^{46}$~erg~s$^{-1}$. Higher luminosity AGNs produce outflows with lower cold gas fractions, with the molecular phase found closer to the nucleus than the ionised one \citep{FioreFeruglio:2017:, FluetschMaiolino:2019:}.

The formation of molecular clumps within outflows reduces the outflow covering fraction as seen from the SMBH. This allows the diffuse hot gas to escape from the bubble and expand further. As it does so, the dense clumps are efficiently compressed by high thermal pressure. The clumps move essentially ballistically and slow down as they climb out of the gravitational potential of the galaxy, while the diffuse gas maintains its velocity. This leads to a small velocity gradient between the cold and hot outflow phases. Such multiphase regions are within the parameter space considered in this work  and can elevate star formation in a galaxy outside the galactic plane. This scenario is consistent with observational evidence of star formation in a significant fraction of outflows \citep{MaiolinoRussell:2017:, GallagherMaiolino:2019:}. Moreover, in the Teacup galaxy (QSO J1430+1339), \citet{VenturiTreister:2023:} found a young stellar population and multiphase outflows with velocity dispersion of $\geq 300$~km~s$^{-1}$. The detected stars coincide with the edges of outflow-blown cavities suggesting positive feedback.

The corollary of this enhancement mode is that the presence of massive molecular clouds is not required for a burst of star formation. Even diffuse, initially almost homogeneous gas can form clumpy, star-forming outflows. If the velocity difference between the clumps and the hot phase is below the threshold limit of cloud dispersal, a starburst is almost guaranteed to occur.

Furthermore, gas-poor ($f_{\rm g} \la 0.1$) systems (e.g. regions where gas has been partially expelled or consumed by stars) can also enhance star formation. The outflows themselves would not necessarily fragment and have velocities above the cloud destruction threshold, the dispersed clouds can mass-load the outflow, increasing the average density of the gas (Fig.~\ref{fig:all_mass}), consistent with findings of \citet{Banda-BarraganZertuche:2019:, GirichidisNaab:2021:}.  As galactic activity is hierarchically clustered in time \citep{HopkinsHernquist:2005:, ZubovasBialopetravicius:2022:}, we expect subsequent AGN-driven outflows to encounter the leftover clumps from previous episodes, still travelling outward with lower relative velocity (i.e. ${<} 400$~km~s$^{-1}$) with respect to the new outflow. The compression by the subsequent activity episodes seems inevitable although it can be delayed by several flow crossing times.

\subsubsection{Cloud compression by fossil outflows}
\label{sec:Fossil outflows}

So far we focused on local effects spanning several dynamic timescales of the clouds. We now address the effects of AGN episode length and variability on cloud compression as it may explain the elusiveness of positive feedback. Typical AGNs flicker between high- and low-Eddington phases, with each cycle lasting approximately  $0.1$~Myr \citep{SchawinskiKoss:2015:, KingNixon:2015:, ZubovasBialopetravicius:2022:}, orders of magnitude shorter than the dynamical time of an outflow. These cycles last for the total duration of an episode, which can be $1\text{ to }10$~Myr \citep{HopkinsHernquist:2005:} comparable to the dynamical time of the molecular clouds. During low-Eddington phases and after the end of the whole episode, outflows persist without obvious nuclear activity and can be seen as fossils. Such coasting outflows are expected to expand for two to three times longer than the driving AGN episode \citep{ZubovasMaskeliunas:2023:}. As the fossil outflow expands in a gas-rich environment, its velocity and pressure gradually decrease, inevitably reaching velocities considered in this work and as such enhancing star formation.  The lower pressure, both thermal and kinematic, of the fossil outflows means that star formation enhancement is less effective and is limited to a narrower velocity range (see Sect.~\ref{section:Fragmentation}). This mode of star formation would be difficult to investigate since outflows may be misclassified as driven by star formation and generally difficult to detect due to low velocity \citep{ZubovasBialopetravicius:2022:}. Future detection and identification of fossil outflows might reveal the level of star formation enhancement in them.

\subsubsection{Supernova-driven outflows}
\label{sec:SN_outflows}

Galaxy-scale outflows are not exclusively driven by AGN, they can also be powered by stellar winds and supernovae. Although typical values of stellar-feedback-driven outflows tend to be lower than those of AGN-driven ones, both samples have overlapping properties. The existence of very high pressure ($5.6 \times 10^7$~K~cm$^{-3}$) fast ($600\text{--}2000$~km~s$^{-1}$) ionised starburst-driven outflows has been predicted for M82 \citep{ChevalierClegg:1985:}. At the other end of the parameter range, \citet{PerrottaCoil:2023:} shows the presence of massive cool $T {\sim} 10^4$~K outflows in starburst galaxies with velocities and mass transfer rates comparable to the sample presented in Sect.~\ref{sec:rel_vel} with the majority of them detected within the central kiloparsec. Similarly, massive cold molecular outflows have been observed in starburst galaxies as well \citep{BolattoWarren:2013:} exhibiting conical geometry \citep{RubinProchaska:2014:, BizyaevChen:2019:}. Finally, as we show in this work, AGN outflows can enhance star formation and so induce supernova-driven outflows.

Additionally, starbursts can form superbubbles comparable in size, velocity and thermal pressure to the AGN-driven fragmenting outflows. For example, the pressure in the NGC 3628 bubble is high ($P/k_{\rm B} \sim 10^{6-8}$~K~cm$^{-3}$) and it is surrounded by slow ($90 \pm 10$~km~s$^{-1}$) molecular gas at 1~kpc from the centre \citep[][]{TsaiMatsushita:2012:}.  It is also well-established that expanding swept-up ISM bubbles can lead to self-propagating star formation via fragmentation of their shells \citep{WhitworthBhattal:1994:}, although this probably requires a threshold luminosity \citep[see e.g.][ and references therein]{WhitworthFrancis:2002:} and the scale of the effect may be limited by interstellar turbulence \citep{NomuraKamaya:2001:}. The observed values of superbubble pressure overlap with the upper portion of the range of pressures of our simulated outflows that lead to enhanced star formation. So our results can be seen as complementary to the self-propagating star formation scenario, but also reveal the possibility of star formation enhancement via compression of pre-existing clouds rather than just those arising from fragmentation of the shell.

\subsection{Kinematics of newly formed stars}\label{sec:Star formation}

Wherever the star formation enhancement occurs (see the previous section), the stars form from significantly perturbed gas. At least part of the perturbation is directional, that is, the cloud is pushed in the direction of the outflow, and the newly formed stars retain some of that momentum. Multiplied by the duration of cloud fragmentation time and the duration of the pre-stellar evolution phase, this can lead to significant displacement of the newly formed stars. Thus, stellar kinematics can be used to identify relatively recent episodes of enhanced star formation.

Stars formed in the `stationary' scenario (Sect. \ref{sec:continuous}) have a low radial velocity ${<} 0.1 v_{\rm out}$ due to long cloud acceleration times. Given that star formation is quenched at outflow velocities above several hundred km~s$^{-1}$, we expect stars formed in this scenario to have radial velocities of order $40$~km~s$^{-1}$ or less. This scenario is compatible with stars being dynamically colder compared to gas \citep{GallagherMaiolino:2019:, OhColless:2022:}. However, the cumulative mass of formed stars is limited by the relatively narrow outflow velocity window capable of enhancing star formation. The window is even narrower at larger radial distances where the molecular clouds are generally smaller and easier to disperse.

Contrary to stationary cloud compression, fragmenting outflows (Sect.~\ref{sec:Clumpy outflows}) result in lower relative velocities between the hot and molecular phases. \citet{GallagherMaiolino:2019:} show that the enhanced star formation `in situ' (i.e. inside the outflow) is common in AGN host galaxies with rates up to $0.3$ of total SFR, and might even dominate in the central kiloparsec. Outflow-formed stars initially have velocities of the outflowing gas\footnote{An alternative hyper velocity star (HVS) launching mechanism - jet launching - is discussed by \citet{DuganBryan:2014:} and \citet{MukherjeeBicknell:2018:}.} and lose their kinetic energy by doing work against gravity. Therefore, the kinematics of the newly formed stars is coupled with the driving pressure, the surrounding gas density and the host galaxy's gravitational potential. If the driving pressure is decreasing (e.g. after the end of an activity episode), the outflow coasts. Contrary to the stationary case, the outflow-formed stars are dynamically hotter as they overtake the outflow. This suggests the stars form in the early stages of the outflow propagation and lose radial momentum rapidly due to gravity. 

In high-luminosity AGNs, in-situ formation from fragmenting outflows is the only viable mechanism of star formation enhancement, since initially stationary clouds are dispersed by the outflow kinematic pressure. The requirement of fragmentation necessitates efficient cooling, which can be negated by the powerful radiation of the AGN itself. As a result, enhanced star formation appears more likely in fossil outflows, where the AGN has already shut down. In this scenario, the newly formed stars have lower transverse velocities than the ones formed from stationary clouds. As pointed out by \citet{ZubovasNayakshin:2013:a}, the velocities can be high enough to escape the bulge and become a part of the galactic halo. Stars with such elongated orbits have been detected in the solar neighbourhood \citep{BelokurovSanders:2020:} and may indicate an early Galactic activity episode. As high-luminosity AGNs can produce massive outflow rates reaching $10^3\text{--}10^4$~M$_{\sun}$~yr${^{-1}}
$ they can easily be the primary source of positive feedback.

As an intermediate case, the kinematics of stars formed via sporadic enhancement in the surroundings of the conical outflows (Sect. \ref{sec:surroundings}) is least certain. The stars retain dynamics of outflow-induced eddies and can have a wide range of radial velocities. The net effect should be an enhancement of the radial anisotropy of stellar velocities in the bulge. However, the properties of these stars potentially resemble both the disc and halo populations and hence makes them very difficult to distinguish and investigate.

\subsection{Comparison with other works}
\label{sec:comparison}

There has been a large number of works, spanning decades, investigating the conditions under which star formation may be enhanced by collisions, outflows or other scenarios of increased external pressure (for a compilation of works see \citealt{DuganGaibler:2017:}, Table 1 and \citealt{Banda-BarraganZertuche:2019:}, Table A1). More recent works explore the destruction threshold values of for cold clouds \citep{LiHopkins:2020:, FarberGronke:2022:}, or even the crushing of multi-cloud systems \citep{VillaresBanda-Barragan:2024:}. Here we compare our results with several works that use outflow velocity and density ranges overlapping with ours.

Our results match those of AGN outflow shocks on Bonnor--Ebbert spheres explored by \citet{DuganGaibler:2017:}. Despite different cloud density profiles compared to our models, and cloud masses comparable to those of cloudlets of our models, they find that outflows of $300$~km~s$^{-1}$, with a temperature of several million~K and particle densities of $0.5, 1.0, 3.0$~cm$^{-3}$, compress the clouds, enhancing star formation. Higher velocity outflows disperse the clouds or produce a negligible mass of stars. They find threshold values for cloud destruction $P/k_{\rm B} = 1.2 \times 10^8$~K~cm$^{-3}$. These values are several times higher compared to our work, due to differences in the adopted survival criteria. If we assume the cloud survival without meeting fragmentation criteria, the cloud destruction threshold pressure is comparable. In our models, there are several surviving cloudlets in the most massive clouds of  $10^5$~M$_{\sun}$, and even the fastest ($400$~km~s$^{-1}$) outflows do not disperse them.

\citet{ZubovasSabulis:2014:} analyse similar systems to the ones in this work; they simulate the impact of slow ($30, 100, 300$~km~s$^{-1}$) outflows on turbulent molecular clouds. They model a spherical turbulent cloud of $\rm M = 10^5$~M$_{\sun}$, with density contrast $\chi = 380$. The cloud density is higher, comparable to lower mass clouds in this work, with turbulent velocity dispersion several times higher.  They find clouds are compressed by external thermal pressure, but outflow velocity has little effect on cloud fragmentation. The fragmentation times are $~{\sim} 1.5$~Myr in $10^5$~K outflows and vary little with outflow velocity. We note the difference in the definition of fragmentation time: we end our simulations when the mass of the formed stars reaches $0.1$~M$_{\rm cloud, init}$, while \citet{ZubovasSabulis:2014:} run theirs until the fraction is ten times higher. Therefore, it is more appropriate to compare our $t_{\rm frag}$ with $t_{\rm sink}$ -- the appearance of first sink (stellar) particles. Despite the differences in density, the qualitative behaviour is similar to our models of the same temperature. If we normalise the timescales by the free-fall time of the clouds, they become similar in both works. They also find no significant differences in fragmentation time in rotating clouds or clouds with high turbulent velocities ${\sim} 10$ ~km~s$^{-1}$.

A recent study by \citet{MandalMukherjee:2024:} explore a similar scenario. They embed the cloud into a hot ambient medium of $10^6$~K and $n=0.1$~cm$^{-3}$ and model propagating wind of $\geq400$~km~s$^{-1}$ and $n=0.01$~cm$^{-3}$, that is, a non-relativistic low-density outflow with moderate velocities. The propagating wind interacts with massive GMCs with particle densities of 20 and 200 cm$^{-3}$. As in our work, the authors include self-gravity and radiative cooling. Despite the differences in the models, authors also lean towards positive, although delayed, feedback even at moderate velocity. Moreover, they show that AGN wind produces multiphase outflows with velocities of $100\text{--}1000$~km~s$^{-1}$ and a wide temperature range ($10^2\text{--}10^7$~K). As a result, positive AGN feedback occurs in two situations. Initially, the low-density wind compresses clouds closer to the AGN, while later, stationary clouds further away are compressed by the dispersed cloud material mixed with the wind, that is, a slow and warm-to-hot outflow. These findings suggest that star formation enhancement is viable as long as the external pressure is maintained.

\citet{CooperBicknell:2009:} investigate cloud compression by star-formation-driven outflows and found similar results to ours. Although they find that star formation is enhanced at a higher velocity of $1200$~km~s$^{-1}$, their simulations have ten times lower outflow density. The ram and thermal pressures are, in fact, comparable to those in our simulations with an outflow temperature of $10^6$~K. They also find a similar morphology -- a fragmented cloud with surviving dense cloudlets embedded in the outflow.

At the opposite extreme, there is little doubt that highly supersonic ($\mathcal M \geq 20$) flows with higher density contrasts compared to our models disperse molecular clouds \citep[e.g.][]{OrlandoPeres:2005:, HopkinsElvis:2010:, ScannapiecoBruggen:2015:}. This suggests that there probably are no other regions of the parameter space where star formation is enhanced.

\subsection{Model caveats} \label{sec:caveats}
\subsubsection{Wind or shock}
\label{sec:wind_shock}

There are two commonly used approaches for simulating the compression of a molecular cloud by an outflow: shock propagation \citep[e.g.][]{PittardParkin:2016:, GoldsmithPittard:2017:} and continuous wind \citep[e.g.][]{Banda-BarraganParkin:2016:, SparrePfrommer:2019:, LiHopkins:2020:}. In the first case, a stationary cloud embedded in the pre-shock medium is struck with the propagating shock that engulfs the cloud. In the continuous wind scenario, a stationary cloud is immersed in uniform gas with velocity and temperature matching those of the post-shock gas. We selected the latter approach due to the simplicity of implementation. However, we note that clouds embedded in a continuous wind experience less compression due to cavities formed in the wake of the cloud and hence have longer evolutionary times comparedk-cloud interaction \citep{GoldsmithPittard:2017:}. We mitigated the discrepancies by ending the simulations at fragmentation time and scaling the results to those of a control simulation. Nevertheless, our approach may underestimate the compressive effect that relatively slow outflows have on interstellar clouds.

\subsubsection{Viability of the cooling function}
\label{sec:shock_cooling}

We estimated the radiative cooling rates via a simplistic cooling function. We assumed an optically thin medium, a constant mean molecular weight, and a constant specific heat ratio for all the gas. Application of a more realistic thermodynamic prescription, especially to the non-equilibrium conditions occurring in the shocks, may alter the results. To justify the validity of our approach we consider the shocks with a velocity of ${\sim} 50$~km~s$^{-1}$, which is the approximate value of cloud shock in $400$~km~s$^{-1}$ outflows and the highest shock velocity in our simulations. Such shocks in dense molecular clouds are of continuous type (C-type) - they have no discontinuities \citep[for review see e.g.][]{DraineMcKee:1993:}. A corollary is that such shocks are non-dissociative, and both molecules and dust survive the passage \citep{DraineRoberge:1983:}. Moreover, in slow to moderate outflows, the dust grains enable the formation of additional coolant molecules, thus preventing a complete shutdown of the cooling channels \citep{HollenbachMcKee:1979:, Neufeld:1990:}. We observed that in our models, the shocked cloud gas is capable of rapid cooling, so ignoring complex shock astrochemistry and optical depth effects is justified.

\subsubsection{Initial conditions and turbulence}

We start our simulations with spherical clouds of uniform density. The initially smooth cloud gas develops stochastic overdensities due to the turbulent velocity field. While such a choice is simple to implement and reduces the parameter space it only partially accounts for the shape of the clouds, as they remain quasi-spherical throughout the control simulations. Realistic clouds are surrounded by diffuse gas envelopes without sharp density or temperature gradients. The presence of such a layer, primarily composed of low-density $n > 10$~cm$^{-3}$ H{\sc I} \citep{HeilesTroland:2003:}, can reduce the overpressure of propagating shock waves. 

Our choice of initial conditions somewhat accounts for the presence of a surrounding gas envelope. The initial density contrast $\chi \simeq 47\text{--}433$ is low enough that it is similar to the contrast between the warm and cold neutral gas phases expected in a real system \citep{HeilesTroland:2003:}. Having such a contrast allows for a realistic development of the internal overdensity structure, so we did not expect major changes in evolution with the inclusion of an additional external diffuse gas layer. 

The geometry of the clouds, however, poses more issues. We expect that the column density of the cloud in the direction parallel to outflow velocity is a key parameter determining how susceptible the cloud is to destruction. If the cloud is strongly non-spherical, its orientation to the outflow becomes important in determining its evolution. We expect to test these issues in future work with a more realistic initial setup.

\section{Summary and conclusions}
\label{sec:Conclusions}

In this work, we have investigated the interaction of hot galactic outflows with isolated molecular clouds and determined the temperature and velocity threshold values for enhancement of cloud fragmentation by the outflow. We carried out SPH simulations of clouds with masses $M_{\rm cloud}=10^{3;4;5}$~M$_{\sun}$ affected by outflows with a constant number density of $1$~cm$^{-3}$, temperatures of $T_{\rm out}=10^{4;5;6}$~K, and velocities $v_{\rm out} = 10, 30, 60, 100, 200, 400$~km~s$^{-1}$ in order to analyse the effect of different thermal and kinematic pressures. We find that slow $\left(v_{\rm out} < 400\, {\rm km~s}^{-1}\right)$ outflows can compress the clouds and induce or enhance ongoing star formation.

The main results of this work are as follows:
\begin{itemize}
    \item We find a single, well-defined region in the cloud and outflow parameter space where star formation is enhanced.
    \item Warm outflows of $10^4$~K with velocities of  $60, 100, 200$~km~s$^{-1}$ enhance star formation. At lower velocities, the outflow has a negligible effect, while faster outflows disperse the massive clouds completely. Star formation in low-mass ($10^3$~M$_{\sun}$) clouds is enhanced at all explored velocities.
    \item At higher outflow temperatures of $10^{5}$~K, thermal pressure compresses the clouds, shortening the fragmentation several times compared to the cooler simulations. The kinematic pressure has little effect on cloud compression in outflows below ${\sim}200$~km~s$^{-1}$, and it disperses massive molecular clouds at higher velocities, similar to the lower temperature outflows.
    \item  In hot outflows of $10^{6}$~K, the clouds are rapidly compressed, and the fragmentation time is reduced by an order of magnitude in the smallest clouds. Kinematic pressure has little effect on cloud evolution, all clouds survive and fragment.
\end{itemize}

We suggest three primary scenarios where star formation enhancement is viable:
\begin{itemize}
    \item Stationary molecular cloud compression in low-powered AGNs where slow outflows develop. This scenario requires both a low velocity and a sustained high outflow temperature to produce positive feedback. It is limited by the availability of cold gas with a low radial velocity.
    \item Sporadic enhancement in shear flow layers surrounding the conical outflows provides a positive feedback mechanism in gas-rich, luminous AGNs. 
    \item Fragmenting multiphase outflows create cold gas with a low velocity (${<} 400$~km~s$^{-1}$) relative to the hot phase. As massive molecular AGN-driven outflows are prevalent, they are the main source of positive feedback.
\end{itemize}

The scenarios are consistent with current observational constraints and with previous works investigating triggered star formation in these disparate domains. Moreover, all three scenarios can occur during and after AGN episode. In the fossil case, positive feedback on star formation occurs without corresponding activity in the nucleus. As a result, enhanced star formation should only weakly correlate with present-day AGN luminosity, except for the central regions of galaxies where the outflow expansion time is short. This temporal discrepancy between cause and effect makes interpretation of observations more difficult. On the other hand, it provides an opportunity to use enhanced star formation as a tool to probe recent (several ayears) nuclear activity histories. 

Peculiar stellar kinematics (see Sect.~\ref{sec:Star formation}) can help with this interpretation as well. In extreme cases, massive AGN-driven molecular outflows can create fountain-like streams of stars. Detection of such stars can be used to investigate former AGN episodes, outflow properties, and dynamics. Additionally, such stars can form peculiar sub-populations in galactic discs. 

Finally, we note that star formation enhancement and outflows from supernovae occur concurrently with AGN-driven outflows. The understanding of gas dynamics in such cycles and the interplay between thermal and kinematic pressure can help in investigations of the difference in the effects of wind, radiation, jet feedback, and/or star formation and can provide a comprehensive view of AGN feedback in general.

\begin{acknowledgements}
We thank the anonymous referee for their insights and improvements to this work. This research was funded by Research Council Lithuania grant no. S-MIP-24-100. The simulations were performed on the supercomputer GALAX of the Center for Physical Sciences and Technology, Lithuania.

\end{acknowledgements}


\bibliographystyle{aa}
\bibliography{literature}

\begin{thebibliography}{110}
\expandafter\ifx\csname natexlab\endcsname\relax\def\natexlab#1{#1}\fi

\bibitem[{{Agertz} {et~al.}(2007){Agertz}, {Moore}, {Stadel}, {Potter},
  {Miniati}, {Read}, {Mayer}, {Gawryszczak}, {Kravtsov}, {Nordlund}, {Pearce},
  {Quilis}, {Rudd}, {Springel}, {Stone}, {Tasker}, {Teyssier}, {Wadsley}, \&
  {Walder}}]{AgertzMoore:2007:}
{Agertz}, O., {Moore}, B., {Stadel}, J., {et~al.} 2007, \mnras, 380, 963

\bibitem[{{Arakawa} {et~al.}(2022){Arakawa}, {Fabian}, {Ferland}, \&
  {Ishibashi}}]{ArakawaFabian:2022:}
{Arakawa}, N., {Fabian}, A.~C., {Ferland}, G.~J., \& {Ishibashi}, W. 2022,
  \mnras, 517, 5069

\bibitem[{{Banda-Barrag{\'a}n} {et~al.}(2016){Banda-Barrag{\'a}n}, {Parkin},
  {Federrath}, {Crocker}, \& {Bicknell}}]{Banda-BarraganParkin:2016:}
{Banda-Barrag{\'a}n}, W.~E., {Parkin}, E.~R., {Federrath}, C., {Crocker},
  R.~M., \& {Bicknell}, G.~V. 2016, \mnras, 455, 1309

\bibitem[{{Banda-Barrag{\'a}n} {et~al.}(2019){Banda-Barrag{\'a}n}, {Zertuche},
  {Federrath}, {Garc{\'\i}a Del Valle}, {Br{\"u}ggen}, \&
  {Wagner}}]{Banda-BarraganZertuche:2019:}
{Banda-Barrag{\'a}n}, W.~E., {Zertuche}, F.~J., {Federrath}, C., {et~al.} 2019,
  \mnras, 486, 4526

\bibitem[{{Belokurov} {et~al.}(2020){Belokurov}, {Sanders}, {Fattahi}, {Smith},
  {Deason}, {Evans}, \& {Grand}}]{BelokurovSanders:2020:}
{Belokurov}, V., {Sanders}, J.~L., {Fattahi}, A., {et~al.} 2020, \mnras, 494,
  3880

\bibitem[{{Bertoldi} \& {McKee}(1992)}]{BertoldiMcKee:1992:}
{Bertoldi}, F. \& {McKee}, C.~F. 1992, \apj, 395, 140

\bibitem[{{Bieri} {et~al.}(2016){Bieri}, {Dubois}, {Silk}, {Mamon}, \&
  {Gaibler}}]{BieriDubois:2016:}
{Bieri}, R., {Dubois}, Y., {Silk}, J., {Mamon}, G.~A., \& {Gaibler}, V. 2016,
  \mnras, 455, 4166

\bibitem[{{Bizyaev} {et~al.}(2019){Bizyaev}, {Chen}, {Shi}, {Riffel}, {Riffel},
  {Diamond-Stanic}, \& {Roy}}]{BizyaevChen:2019:}
{Bizyaev}, D., {Chen}, Y.-M., {Shi}, Y., {et~al.} 2019, \apj, 882, 145

\bibitem[{{Bolatto} {et~al.}(2013){Bolatto}, {Warren}, {Leroy}, {Walter},
  {Veilleux}, {Ostriker}, {Ott}, {Zwaan}, {Fisher}, {Weiss}, {Rosolowsky}, \&
  {Hodge}}]{BolattoWarren:2013:}
{Bolatto}, A.~D., {Warren}, S.~R., {Leroy}, A.~K., {et~al.} 2013, \nat, 499,
  450

\bibitem[{{Chevalier} \& {Clegg}(1985)}]{ChevalierClegg:1985:}
{Chevalier}, R.~A. \& {Clegg}, A.~W. 1985, \nat, 317, 44

\bibitem[{{Chevance} {et~al.}(2020){Chevance}, {Kruijssen}, {Vazquez-Semadeni},
  {Nakamura}, {Klessen}, {Ballesteros-Paredes}, {Inutsuka}, {Adamo}, \&
  {Hennebelle}}]{ChevanceKruijssen:2020:}
{Chevance}, M., {Kruijssen}, J.~M.~D., {Vazquez-Semadeni}, E., {et~al.} 2020,
  \ssr, 216, 50

\bibitem[{{Cooper} {et~al.}(2009){Cooper}, {Bicknell}, {Sutherland}, \&
  {Bland-Hawthorn}}]{CooperBicknell:2009:}
{Cooper}, J.~L., {Bicknell}, G.~V., {Sutherland}, R.~S., \& {Bland-Hawthorn},
  J. 2009, \apj, 703, 330

\bibitem[{{Costa} {et~al.}(2020){Costa}, {Pakmor}, \&
  {Springel}}]{CostaPakmor:2020:}
{Costa}, T., {Pakmor}, R., \& {Springel}, V. 2020, \mnras, 497, 5229

\bibitem[{{Cowie} \& {McKee}(1977)}]{CowieMcKee:1977:}
{Cowie}, L.~L. \& {McKee}, C.~F. 1977, \apj, 211, 135

\bibitem[{{Cresci} {et~al.}(2015){Cresci}, {Mainieri}, {Brusa}, {Marconi},
  {Perna}, {Mannucci}, {Piconcelli}, {Maiolino}, {Feruglio}, {Fiore},
  {Bongiorno}, {Lanzuisi}, {Merloni}, {Schramm}, {Silverman}, \&
  {Civano}}]{CresciMainieri:2015:}
{Cresci}, G., {Mainieri}, V., {Brusa}, M., {et~al.} 2015, \apj, 799, 82

\bibitem[{{Dahmer-Hahn} {et~al.}(2022){Dahmer-Hahn}, {Riffel},
  {Rodr{\'\i}guez-Ardila}, {Riffel}, {Storchi-Bergmann}, {Marinello}, {Davies},
  {Burtscher}, {Ruschel-Dutra}, \& {Rosario}}]{Dahmer-HahnRiffel:2022:}
{Dahmer-Hahn}, L.~G., {Riffel}, R., {Rodr{\'\i}guez-Ardila}, A., {et~al.} 2022,
  \mnras, 509, 4653

\bibitem[{{Dehnen} \& {Aly}(2012)}]{DehnenAly:2012:}
{Dehnen}, W. \& {Aly}, H. 2012, \mnras, 425, 1068

\bibitem[{{Draine} \& {McKee}(1993)}]{DraineMcKee:1993:}
{Draine}, B.~T. \& {McKee}, C.~F. 1993, \araa, 31, 373

\bibitem[{{Draine} {et~al.}(1983){Draine}, {Roberge}, \&
  {Dalgarno}}]{DraineRoberge:1983:}
{Draine}, B.~T., {Roberge}, W.~G., \& {Dalgarno}, A. 1983, \apj, 264, 485

\bibitem[{{Dubinski} {et~al.}(1995){Dubinski}, {Narayan}, \&
  {Phillips}}]{DubinskiNarayan:1995:}
{Dubinski}, J., {Narayan}, R., \& {Phillips}, T.~G. 1995, \apj, 448, 226

\bibitem[{{Dugan} {et~al.}(2014){Dugan}, {Bryan}, {Gaibler}, {Silk}, \&
  {Haas}}]{DuganBryan:2014:}
{Dugan}, Z., {Bryan}, S., {Gaibler}, V., {Silk}, J., \& {Haas}, M. 2014, \apj,
  796, 113

\bibitem[{{Dugan} {et~al.}(2017){Dugan}, {Gaibler}, {Bieri}, {Silk}, \&
  {Rahman}}]{DuganGaibler:2017:}
{Dugan}, Z., {Gaibler}, V., {Bieri}, R., {Silk}, J., \& {Rahman}, M. 2017,
  \apj, 839, 103

\bibitem[{{Fabian}(2012)}]{Fabian:2012:}
{Fabian}, A.~C. 2012, \araa, 50, 455

\bibitem[{{Farber} \& {Gronke}(2022)}]{FarberGronke:2022:}
{Farber}, R.~J. \& {Gronke}, M. 2022, \mnras, 510, 551

\bibitem[{{Faucher-Gigu{\`e}re} \&
  {Quataert}(2012)}]{Faucher-GiguereQuataert:2012:}
{Faucher-Gigu{\`e}re}, C.-A. \& {Quataert}, E. 2012, \mnras, 425, 605

\bibitem[{{Federrath} {et~al.}(2008){Federrath}, {Klessen}, \&
  {Schmidt}}]{FederrathKlessen:2008:}
{Federrath}, C., {Klessen}, R.~S., \& {Schmidt}, W. 2008, \apjl, 688, L79

\bibitem[{{Ferrarese} \& {Merritt}(2000)}]{FerrareseMerritt:2000:}
{Ferrarese}, L. \& {Merritt}, D. 2000, \apjl, 539, L9

\bibitem[{{Fiore} {et~al.}(2017){Fiore}, {Feruglio}, {Shankar}, {Bischetti},
  {Bongiorno}, {Brusa}, {Carniani}, {Cicone}, {Duras}, {Lamastra}, {Mainieri},
  {Marconi}, {Menci}, {Maiolino}, {Piconcelli}, {Vietri}, \&
  {Zappacosta}}]{FioreFeruglio:2017:}
{Fiore}, F., {Feruglio}, C., {Shankar}, F., {et~al.} 2017, \aap, 601, A143

\bibitem[{{Fluetsch} {et~al.}(2021){Fluetsch}, {Maiolino}, {Carniani},
  {Arribas}, {Belfiore}, {Bellocchi}, {Cazzoli}, {Cicone}, {Cresci}, {Fabian},
  {Gallagher}, {Ishibashi}, {Mannucci}, {Marconi}, {Perna}, {Sturm}, \&
  {Venturi}}]{FluetschMaiolino:2021:}
{Fluetsch}, A., {Maiolino}, R., {Carniani}, S., {et~al.} 2021, \mnras, 505,
  5753

\bibitem[{{Fluetsch} {et~al.}(2019){Fluetsch}, {Maiolino}, {Carniani},
  {Marconi}, {Cicone}, {Bourne}, {Costa}, {Fabian}, {Ishibashi}, \&
  {Venturi}}]{FluetschMaiolino:2019:}
{Fluetsch}, A., {Maiolino}, R., {Carniani}, S., {et~al.} 2019, \mnras, 483,
  4586

\bibitem[{{Gallagher} {et~al.}(2019){Gallagher}, {Maiolino}, {Belfiore},
  {Drory}, {Riffel}, \& {Riffel}}]{GallagherMaiolino:2019:}
{Gallagher}, R., {Maiolino}, R., {Belfiore}, F., {et~al.} 2019, \mnras, 485,
  3409

\bibitem[{{Ginsburg} {et~al.}(2013){Ginsburg}, {Federrath}, \&
  {Darling}}]{GinsburgFederrath:2013:}
{Ginsburg}, A., {Federrath}, C., \& {Darling}, J. 2013, \apj, 779, 50

\bibitem[{{Girichidis} {et~al.}(2021){Girichidis}, {Naab}, {Walch}, \&
  {Berlok}}]{GirichidisNaab:2021:}
{Girichidis}, P., {Naab}, T., {Walch}, S., \& {Berlok}, T. 2021, \mnras, 505,
  1083

\bibitem[{{Goldsmith} \& {Pittard}(2017)}]{GoldsmithPittard:2017:}
{Goldsmith}, K.~J.~A. \& {Pittard}, J.~M. 2017, \mnras, 470, 2427

\bibitem[{{Heiles} \& {Troland}(2003)}]{HeilesTroland:2003:}
{Heiles}, C. \& {Troland}, T.~H. 2003, \apj, 586, 1067

\bibitem[{{Ho}(2008)}]{Ho:2008:}
{Ho}, L.~C. 2008, \araa, 46, 475

\bibitem[{{Hollenbach} \& {McKee}(1979)}]{HollenbachMcKee:1979:}
{Hollenbach}, D. \& {McKee}, C.~F. 1979, \apjs, 41, 555

\bibitem[{{Hopkins}(2013)}]{Hopkins:2013:}
{Hopkins}, P.~F. 2013, \mnras, 428, 2840

\bibitem[{{Hopkins} \& {Elvis}(2010)}]{HopkinsElvis:2010:}
{Hopkins}, P.~F. \& {Elvis}, M. 2010, \mnras, 401, 7

\bibitem[{{Hopkins} {et~al.}(2005){Hopkins}, {Hernquist}, {Martini}, {Cox},
  {Robertson}, {Di Matteo}, \& {Springel}}]{HopkinsHernquist:2005:}
{Hopkins}, P.~F., {Hernquist}, L., {Martini}, P., {et~al.} 2005, \apjl, 625,
  L71

\bibitem[{Hunter(2007)}]{Hunter:2007}
Hunter, J.~D. 2007, Computing in Science \& Engineering, 9, 90

\bibitem[{{Ilha} {et~al.}(2019){Ilha}, {Riffel}, {Schimoia},
  {Storchi-Bergmann}, {Rembold}, {Riffel}, {Wylezalek}, {Shi}, {da Costa},
  {Machado}, {Law}, {Bizyaev}, {Mallmann}, {Nascimento}, {Maia}, \&
  {Cirolini}}]{IlhaRiffel:2019:}
{Ilha}, G.~S., {Riffel}, R.~A., {Schimoia}, J.~S., {et~al.} 2019, \mnras, 484,
  252

\bibitem[{{Inoue} \& {Inutsuka}(2008)}]{InoueInutsuka:2008:}
{Inoue}, T. \& {Inutsuka}, S.-i. 2008, \apj, 687, 303

\bibitem[{{Kakiuchi} {et~al.}(2024){Kakiuchi}, {Suzuki}, {Inutsuka}, {Inoue},
  \& {Shimoda}}]{KakiuchiSuzuki:2024:}
{Kakiuchi}, K., {Suzuki}, T.~K., {Inutsuka}, S.-i., {Inoue}, T., \& {Shimoda},
  J. 2024, \apj, 966, 230

\bibitem[{{Kennicutt} \& {Evans}(2012)}]{KennicuttEvans:2012:}
{Kennicutt}, R.~C. \& {Evans}, N.~J. 2012, \araa, 50, 531

\bibitem[{{Khullar} {et~al.}(2019){Khullar}, {Krumholz}, {Federrath}, \&
  {Cunningham}}]{KhullarKrumholz:2019:}
{Khullar}, S., {Krumholz}, M.~R., {Federrath}, C., \& {Cunningham}, A.~J. 2019,
  \mnras, 488, 1407

\bibitem[{{King} \& {Nixon}(2015)}]{KingNixon:2015:}
{King}, A. \& {Nixon}, C. 2015, \mnras, 453, L46

\bibitem[{{King} \& {Pounds}(2015)}]{KingPounds:2015:}
{King}, A. \& {Pounds}, K. 2015, \araa, 53, 115

\bibitem[{{Klein} {et~al.}(1994){Klein}, {McKee}, \&
  {Colella}}]{KleinMcKee:1994:}
{Klein}, R.~I., {McKee}, C.~F., \& {Colella}, P. 1994, \apj, 420, 213

\bibitem[{{Kormendy} \& {Ho}(2013)}]{KormendyHo:2013:}
{Kormendy}, J. \& {Ho}, L.~C. 2013, \araa, 51, 511

\bibitem[{{Koyama} \& {Inutsuka}(2000)}]{KoyamaInutsuka:2000:}
{Koyama}, H. \& {Inutsuka}, S.-I. 2000, \apj, 532, 980

\bibitem[{{Koyama} \& {Inutsuka}(2002)}]{KoyamaInutsuka:2002:}
{Koyama}, H. \& {Inutsuka}, S.-i. 2002, \apjl, 564, L97

\bibitem[{{Laha} {et~al.}(2021){Laha}, {Reynolds}, {Reeves}, {Kriss},
  {Guainazzi}, {Smith}, {Veilleux}, \& {Proga}}]{LahaReynolds:2021:}
{Laha}, S., {Reynolds}, C.~S., {Reeves}, J., {et~al.} 2021, Nature Astronomy,
  5, 13

\bibitem[{{Landau} \& {Lifshitz}(1959)}]{LandauLifshitz:1959:}
{Landau}, L.~D. \& {Lifshitz}, E.~M. 1959, {Fluid mechanics}

\bibitem[{{Larson}(1981)}]{Larson:1981:}
{Larson}, R.~B. 1981, \mnras, 194, 809

\bibitem[{{Li} {et~al.}(2020){Li}, {Hopkins}, {Squire}, \&
  {Hummels}}]{LiHopkins:2020:}
{Li}, Z., {Hopkins}, P.~F., {Squire}, J., \& {Hummels}, C. 2020, \mnras, 492,
  1841

\bibitem[{{Lutz} {et~al.}(2020){Lutz}, {Sturm}, {Janssen}, {Veilleux}, {Aalto},
  {Cicone}, {Contursi}, {Davies}, {Feruglio}, {Fischer}, {Fluetsch},
  {Garcia-Burillo}, {Genzel}, {Gonz{\'a}lez-Alfonso}, {Graci{\'a}-Carpio},
  {Herrera-Camus}, {Maiolino}, {Schruba}, {Shimizu}, {Sternberg}, {Tacconi}, \&
  {Wei{\ss}}}]{LutzSturm:2020:}
{Lutz}, D., {Sturm}, E., {Janssen}, A., {et~al.} 2020, \aap, 633, A134

\bibitem[{{Maiolino} {et~al.}(2017){Maiolino}, {Russell}, {Fabian}, {Carniani},
  {Gallagher}, {Cazzoli}, {Arribas}, {Belfiore}, {Bellocchi}, {Colina},
  {Cresci}, {Ishibashi}, {Marconi}, {Mannucci}, {Oliva}, \&
  {Sturm}}]{MaiolinoRussell:2017:}
{Maiolino}, R., {Russell}, H.~R., {Fabian}, A.~C., {et~al.} 2017, \nat, 544,
  202

\bibitem[{{Mandal} {et~al.}(2024){Mandal}, {Mukherjee}, {Federrath},
  {Bicknell}, {Nesvadba}, \& {Mignone}}]{MandalMukherjee:2024:}
{Mandal}, A., {Mukherjee}, D., {Federrath}, C., {et~al.} 2024, \mnras, 531,
  2079

\bibitem[{{Mercedes-Feliz} {et~al.}(2023){Mercedes-Feliz},
  {Angl{\'e}s-Alc{\'a}zar}, {Hayward}, {Cochrane}, {Terrazas}, {Wellons},
  {Richings}, {Faucher-Gigu{\`e}re}, {Moreno}, {Su}, {Hopkins}, {Quataert}, \&
  {Kere{\v{s}}}}]{Mercedes-FelizAngles-Alcazar:2023:}
{Mercedes-Feliz}, J., {Angl{\'e}s-Alc{\'a}zar}, D., {Hayward}, C.~C., {et~al.}
  2023, \mnras, 524, 3446

\bibitem[{{Miville-Desch{\^e}nes} {et~al.}(2017){Miville-Desch{\^e}nes},
  {Murray}, \& {Lee}}]{Miville-DeschenesMurray:2017:}
{Miville-Desch{\^e}nes}, M.-A., {Murray}, N., \& {Lee}, E.~J. 2017, \apj, 834,
  57

\bibitem[{{Mukherjee} {et~al.}(2018){Mukherjee}, {Bicknell}, {Wagner},
  {Sutherland}, \& {Silk}}]{MukherjeeBicknell:2018:}
{Mukherjee}, D., {Bicknell}, G.~V., {Wagner}, A.~Y., {Sutherland}, R.~S., \&
  {Silk}, J. 2018, \mnras, 479, 5544

\bibitem[{{Murray}(2011)}]{Murray:2011:}
{Murray}, N. 2011, \apj, 729, 133

\bibitem[{{Nayakshin} \& {Zubovas}(2012)}]{NayakshinZubovas:2012:}
{Nayakshin}, S. \& {Zubovas}, K. 2012, \mnras, 427, 372

\bibitem[{{Neufeld}(1990)}]{Neufeld:1990:}
{Neufeld}, D.~A. 1990, in Molecular Astrophysics, 374

\bibitem[{{Nevin} {et~al.}(2018){Nevin}, {Comerford}, {M{\"u}ller-S{\'a}nchez},
  {Barrows}, \& {Cooper}}]{NevinComerford:2018:}
{Nevin}, R., {Comerford}, J.~M., {M{\"u}ller-S{\'a}nchez}, F., {Barrows}, R.,
  \& {Cooper}, M.~C. 2018, \mnras, 473, 2160

\bibitem[{{Nomura} \& {Kamaya}(2001)}]{NomuraKamaya:2001:}
{Nomura}, H. \& {Kamaya}, H. 2001, \aj, 121, 1024

\bibitem[{{Oh} {et~al.}(2022){Oh}, {Colless}, {D'Eugenio}, {Croom}, {Cortese},
  {Groves}, {Kewley}, {van de Sande}, {Zovaro}, {Varidel}, {Barsanti},
  {Bland-Hawthorn}, {Brough}, {Bryant}, {Casura}, {Lawrence}, {Lorente},
  {Medling}, {Owers}, \& {Yi}}]{OhColless:2022:}
{Oh}, S., {Colless}, M., {D'Eugenio}, F., {et~al.} 2022, \mnras, 512, 1765

\bibitem[{{Orlando} {et~al.}(2005){Orlando}, {Peres}, {Reale}, {Bocchino},
  {Rosner}, {Plewa}, \& {Siegel}}]{OrlandoPeres:2005:}
{Orlando}, S., {Peres}, G., {Reale}, F., {et~al.} 2005, \aap, 444, 505

\bibitem[{{Perrotta} {et~al.}(2023){Perrotta}, {Coil}, {Rupke}, {Tremonti},
  {Davis}, {Diamond-Stanic}, {Geach}, {Hickox}, {Moustakas}, {Rudnick}, {Sell},
  {Swiggum}, \& {Whalen}}]{PerrottaCoil:2023:}
{Perrotta}, S., {Coil}, A.~L., {Rupke}, D. S.~N., {et~al.} 2023, \apj, 949, 9

\bibitem[{{Pittard} {et~al.}(2010){Pittard}, {Hartquist}, \&
  {Falle}}]{PittardHartquist:2010:}
{Pittard}, J.~M., {Hartquist}, T.~W., \& {Falle}, S.~A.~E.~G. 2010, \mnras,
  405, 821

\bibitem[{{Pittard} \& {Parkin}(2016)}]{PittardParkin:2016:}
{Pittard}, J.~M. \& {Parkin}, E.~R. 2016, \mnras, 457, 4470

\bibitem[{{Poludnenko} {et~al.}(2002){Poludnenko}, {Frank}, \&
  {Blackman}}]{PoludnenkoFrank:2002:}
{Poludnenko}, A.~Y., {Frank}, A., \& {Blackman}, E.~G. 2002, \apj, 576, 832

\bibitem[{{Richings} \&
  {Faucher-Gigu{\`e}re}(2018)}]{RichingsFaucher-Giguere:2018:}
{Richings}, A.~J. \& {Faucher-Gigu{\`e}re}, C.-A. 2018, \mnras, 474, 3673

\bibitem[{{Rubin} {et~al.}(2014){Rubin}, {Prochaska}, {Koo}, {Phillips},
  {Martin}, \& {Winstrom}}]{RubinProchaska:2014:}
{Rubin}, K. H.~R., {Prochaska}, J.~X., {Koo}, D.~C., {et~al.} 2014, \apj, 794,
  156

\bibitem[{{Scannapieco}(2017)}]{Scannapieco:2017:}
{Scannapieco}, E. 2017, \apj, 837, 28

\bibitem[{{Scannapieco} \& {Br{\"u}ggen}(2015)}]{ScannapiecoBruggen:2015:}
{Scannapieco}, E. \& {Br{\"u}ggen}, M. 2015, \apj, 805, 158

\bibitem[{{Schawinski} {et~al.}(2015){Schawinski}, {Koss}, {Berney}, \&
  {Sartori}}]{SchawinskiKoss:2015:}
{Schawinski}, K., {Koss}, M., {Berney}, S., \& {Sartori}, L.~F. 2015, \mnras,
  451, 2517

\bibitem[{{Schawinski} {et~al.}(2014){Schawinski}, {Urry}, {Simmons},
  {Fortson}, {Kaviraj}, {Keel}, {Lintott}, {Masters}, {Nichol}, {Sarzi},
  {Skibba}, {Treister}, {Willett}, {Wong}, \& {Yi}}]{SchawinskiUrry:2014:}
{Schawinski}, K., {Urry}, C.~M., {Simmons}, B.~D., {et~al.} 2014, \mnras, 440,
  889

\bibitem[{Schubert {et~al.}(2017)Schubert, Sander, Ester, Kriegel, \&
  Xu}]{10.1145/3068335}
Schubert, E., Sander, J., Ester, M., Kriegel, H.~P., \& Xu, X. 2017, ACM Trans.
  Database Syst., 42

\bibitem[{{Silk}(2013)}]{Silk:2013:}
{Silk}, J. 2013, \apj, 772, 112

\bibitem[{{Sparre} {et~al.}(2019){Sparre}, {Pfrommer}, \&
  {Vogelsberger}}]{SparrePfrommer:2019:}
{Sparre}, M., {Pfrommer}, C., \& {Vogelsberger}, M. 2019, \mnras, 482, 5401

\bibitem[{{Spitzer}(1978)}]{Spitzer:1978:}
{Spitzer}, L. 1978, {Physical processes in the interstellar medium}

\bibitem[{{Spreiter} {et~al.}(1966){Spreiter}, {Summers}, \&
  {Alksne}}]{SpreiterSummers:1966:}
{Spreiter}, J.~R., {Summers}, A.~L., \& {Alksne}, A.~Y. 1966, \planss, 14,
  223,IN1,251

\bibitem[{{Springel} {et~al.}(2021){Springel}, {Pakmor}, {Zier}, \&
  {Reinecke}}]{SpringelPakmor:2021:}
{Springel}, V., {Pakmor}, R., {Zier}, O., \& {Reinecke}, M. 2021, \mnras, 506,
  2871

\bibitem[{{Sun} {et~al.}(2018){Sun}, {Leroy}, {Schruba}, {Rosolowsky},
  {Hughes}, {Kruijssen}, {Meidt}, {Schinnerer}, {Blanc}, {Bigiel}, {Bolatto},
  {Chevance}, {Groves}, {Herrera}, {Hygate}, {Pety}, {Querejeta}, {Usero}, \&
  {Utomo}}]{SunLeroy:2018:}
{Sun}, J., {Leroy}, A.~K., {Schruba}, A., {et~al.} 2018, \apj, 860, 172

\bibitem[{{Sutherland} \& {Dopita}(1993)}]{SutherlandDopita:1993:}
{Sutherland}, R.~S. \& {Dopita}, M.~A. 1993, \apjs, 88, 253

\bibitem[{{Thompson} {et~al.}(2016){Thompson}, {Quataert}, {Zhang}, \&
  {Weinberg}}]{ThompsonQuataert:2016:}
{Thompson}, T.~A., {Quataert}, E., {Zhang}, D., \& {Weinberg}, D.~H. 2016,
  \mnras, 455, 1830

\bibitem[{{Tsai} {et~al.}(2012){Tsai}, {Matsushita}, {Kong}, {Matsumoto}, \&
  {Kohno}}]{TsaiMatsushita:2012:}
{Tsai}, A.-L., {Matsushita}, S., {Kong}, A. K.~H., {Matsumoto}, H., \& {Kohno},
  K. 2012, \apj, 752, 38

\bibitem[{{Valentini} {et~al.}(2017){Valentini}, {Murante}, {Borgani},
  {Monaco}, {Bressan}, \& {Beck}}]{ValentiniMurante:2017:}
{Valentini}, M., {Murante}, G., {Borgani}, S., {et~al.} 2017, \mnras, 470, 3167

\bibitem[{{Veilleux} {et~al.}(2020){Veilleux}, {Maiolino}, {Bolatto}, \&
  {Aalto}}]{VeilleuxMaiolino:2020:}
{Veilleux}, S., {Maiolino}, R., {Bolatto}, A.~D., \& {Aalto}, S. 2020, \aapr,
  28, 2

\bibitem[{{Venturi} {et~al.}(2023){Venturi}, {Treister}, {Finlez}, {D'Ago},
  {Bauer}, {Harrison}, {Ramos Almeida}, {Revalski}, {Ricci}, {Sartori},
  {Girdhar}, {Keel}, \& {Tub{\'\i}n}}]{VenturiTreister:2023:}
{Venturi}, G., {Treister}, E., {Finlez}, C., {et~al.} 2023, \aap, 678, A127

\bibitem[{{Villares} {et~al.}(2024){Villares}, {Banda-Barrag{\'a}n}, \&
  {Rojas}}]{VillaresBanda-Barragan:2024:}
{Villares}, A.~S., {Banda-Barrag{\'a}n}, W.~E., \& {Rojas}, C. 2024, \mnras,
  submitted

\bibitem[{{Ward} {et~al.}(2024){Ward}, {Costa}, {Harrison}, \&
  {Mainieri}}]{WardCosta:2024:}
{Ward}, S.~R., {Costa}, T., {Harrison}, C.~M., \& {Mainieri}, V. 2024, \mnras,
  in press

\bibitem[{{Weaver} {et~al.}(1977){Weaver}, {McCray}, {Castor}, {Shapiro}, \&
  {Moore}}]{WeaverMcCray:1977:}
{Weaver}, R., {McCray}, R., {Castor}, J., {Shapiro}, P., \& {Moore}, R. 1977,
  \apj, 218, 377

\bibitem[{{Whitworth} {et~al.}(1994){Whitworth}, {Bhattal}, {Chapman},
  {Disney}, \& {Turner}}]{WhitworthBhattal:1994:}
{Whitworth}, A.~P., {Bhattal}, A.~S., {Chapman}, S.~J., {Disney}, M.~J., \&
  {Turner}, J.~A. 1994, \mnras, 268, 291

\bibitem[{{Whitworth} \& {Francis}(2002)}]{WhitworthFrancis:2002:}
{Whitworth}, A.~P. \& {Francis}, N. 2002, \mnras, 329, 641

\bibitem[{{Wurster} \& {Thacker}(2013)}]{WursterThacker:2013:}
{Wurster}, J. \& {Thacker}, R.~J. 2013, \mnras, 431, 2513

\bibitem[{{Zhou} {et~al.}(2021){Zhou}, {Williams}, {Ramaprabhu}, {Groom},
  {Thornber}, {Hillier}, {Mostert}, {Rollin}, {Balachandar}, {Powell},
  {Mahalov}, \& {Attal}}]{ZhouWilliams:2021:}
{Zhou}, Y., {Williams}, R. J.~R., {Ramaprabhu}, P., {et~al.} 2021, Physica D
  Nonlinear Phenomena, 423, 132838

\bibitem[{{Zubovas} {et~al.}(2022){Zubovas}, {Bialopetravi{\v{c}}ius}, \&
  {Kazlauskait{\.{e}}}}]{ZubovasBialopetravicius:2022:}
{Zubovas}, K., {Bialopetravi{\v{c}}ius}, J., \& {Kazlauskait{\.{e}}}, M. 2022,
  \mnras, 515, 1705

\bibitem[{{Zubovas} \& {King}(2012{\natexlab{a}})}]{ZubovasKing:2012:b}
{Zubovas}, K. \& {King}, A. 2012{\natexlab{a}}, \apjl, 745, L34

\bibitem[{{Zubovas} \& {King}(2016)}]{ZubovasKing:2016:}
{Zubovas}, K. \& {King}, A. 2016, \mnras, 462, 4055

\bibitem[{{Zubovas} \& {King}(2012{\natexlab{b}})}]{ZubovasKing:2012:a}
{Zubovas}, K. \& {King}, A.~R. 2012{\natexlab{b}}, \mnras, 426, 2751

\bibitem[{{Zubovas} \& {King}(2014)}]{ZubovasKing:2014:}
{Zubovas}, K. \& {King}, A.~R. 2014, \mnras, 439, 400

\bibitem[{{Zubovas} \& {King}(2019)}]{ZubovasKing:2019:}
{Zubovas}, K. \& {King}, A.~R. 2019, General Relativity and Gravitation, 51, 65

\bibitem[{{Zubovas} \& {Maskeli{\={u}}nas}(2023)}]{ZubovasMaskeliunas:2023:}
{Zubovas}, K. \& {Maskeli{\={u}}nas}, G. 2023, \mnras, 524, 4819

\bibitem[{{Zubovas} \& {Nardini}(2020)}]{ZubovasNardini:2020:}
{Zubovas}, K. \& {Nardini}, E. 2020, \mnras, 498, 3633

\bibitem[{{Zubovas} {et~al.}(2013{\natexlab{a}}){Zubovas}, {Nayakshin}, {King},
  \& {Wilkinson}}]{ZubovasNayakshin:2013:a}
{Zubovas}, K., {Nayakshin}, S., {King}, A., \& {Wilkinson}, M.
  2013{\natexlab{a}}, \mnras, 433, 3079

\bibitem[{{Zubovas} {et~al.}(2013{\natexlab{b}}){Zubovas}, {Nayakshin},
  {Sazonov}, \& {Sunyaev}}]{ZubovasNayakshin:2013:b}
{Zubovas}, K., {Nayakshin}, S., {Sazonov}, S., \& {Sunyaev}, R.
  2013{\natexlab{b}}, \mnras, 431, 793

\bibitem[{{Zubovas} {et~al.}(2014){Zubovas}, {Sabulis}, \&
  {Naujalis}}]{ZubovasSabulis:2014:}
{Zubovas}, K., {Sabulis}, K., \& {Naujalis}, R. 2014, \mnras, 442, 2837

\end{thebibliography}


\onecolumn 
\begin{appendix} 
\section{Model list} \label{ap:parameter_table}
In Table~\ref{table:models_all} we present a summary of all the simulations performed in this work. The first three columns show the outflow and cloud parameters that we vary, followed by four derived parameters: density contrast, Mach number, cloud crushing time (Eq.~\ref{eq:tcc}) and the cloud acceleration time $t_{\rm acc} \approx \chi^{1/2} t_{\rm cc}$. The following two columns show the dimensions of the simulation volumes (elongated `boxes') used. While we freely choose the size of the box perpendicular to the flow ($y = z$), the code we used limits the choice of length ratios to powers of two ($x = 2^n\times y$). We note the lengths of some boxes are comparable to the sizes of bulges in small galaxies. The last two columns show the main results of this work -- fragmentation times, in both non-normalised and normalised (Eq. \ref{eq:t_norm}) forms. Missing entries here indicate the models where clouds were dispersed without reaching fragmentation criteria (see Sect.~\ref{sec:SF}).
\begin{table}[h!]
\protect\caption{Summary of the simulations.}
\centering{}%
\label{table:models_all}
\scalebox{0.7}{
\begin{tabular}{cccccccccll}
\hline\hline
$M_{\rm cl}$ [M$_\sun$] & $T_{\rm out}$ [K] & $v_{\rm out}$ [km~s$^{-1}$] & $\chi$ & $\mathcal{M}$ & $t_{\rm cc}$ [Myr]& $t_{\rm acc}$ [Myr] & Box$_{y,z}$ [pc] & Box$_x$ [pc] &$t_{\rm frag}$ [Myr]&$t_{\rm norm}$\\
\hline
$10^3$ & $10^4$ & 0  & 433 & -  & -  & - & 80 & 80   &2.43&1\\
$10^3$ & $10^4$ & 10  & 433 & 0.68  & 6.0  & 133.1 & 80 & 160   &2.17&0.89\\
$10^3$ & $10^4$ & 30  & 433 & 2.03  & 2.0  & 44.4  & 80 & 160   &1.65&0.68\\
$10^3$ & $10^4$ & 60  & 433 & 4.06  & 1.0  & 22.2  & 80 & 640   &1.51&0.62\\
$10^3$ & $10^4$ & 100 & 433 & 6.77  & 0.6  & 13.3  & 80 & 640   &1.63&0.67\\
$10^3$ & $10^4$ & 200 & 433 & 13.53 & 0.3  & 6.7   & 80 & 1280   &-  &0.48\\
$10^3$ & $10^4$ & 400 & 433 & 27.07 & 0.1  & 3.3   & 80 & 1280  &-  &0.57\\
$10^3$ & $10^5$ & 10  & 433 & 0.21  & 6.0  & 133.1 & 80 & 160   &0.78&0.32\\
$10^3$ & $10^5$ & 30  & 433 & 0.63  & 2.0  & 44.4  & 80 & 160   &0.78&0.32\\
$10^3$ & $10^5$ & 60  & 433 & 1.28  & 1.0  & 22.2  & 80 & 320   &0.71&0.29\\
$10^3$ & $10^5$ & 100 & 433 & 2.14  & 0.6  & 13.3  & 80 & 320   &0.63&0.26\\
$10^3$ & $10^5$ & 200 & 433 & 4.28  & 0.3  & 6.7   & 80 & 640   &0.78&0.32\\
$10^3$ & $10^5$ & 400 & 433 & 8.56  & 0.1  & 3.3   & 80 & 1280  &-  &0.33\\
$10^3$ & $10^6$ & 10  & 433 & 0.07  & 6.0  & 133.1 & 80 & 160   &0.29&0.12\\
$10^3$ & $10^6$ & 30  & 433 & 0.20  & 2.0  & 44.4  & 80 & 160   &0.29&0.12\\
$10^3$ & $10^6$ & 60  & 433 & 0.41  & 1.0  & 22.2  & 80 & 160   &0.29&0.12\\
$10^3$ & $10^6$ & 100 & 433 & 0.68  & 0.6  & 13.3  & 80 & 160   &0.29&0.12\\
$10^3$ & $10^6$ & 200 & 433 & 1.35  & 0.3  & 6.7   & 80 & 160   &0.24&0.10\\
$10^3$ & $10^6$ & 400 & 433 & 2.71  & 0.1  & 3.3   & 80 & 640   &0.17&0.07\\
$10^4$ & $10^4$ & 0  & 143 & -  & - & - & 100 & 100   &5.50  &1\\
$10^4$ & $10^4$ & 10  & 143 & 0.68  & 10.5 & 127.9 & 100 & 200   &5.47&0.99\\
$10^4$ & $10^4$ & 30  & 143 & 2.03  & 3.5  & 42.6  & 100 & 200   &4.80&0.87\\
$10^4$ & $10^4$ & 60  & 143 & 4.06  & 1.8  & 21.3  & 100 & 400   &4.33&0.78\\
$10^4$ & $10^4$ & 100 & 143 & 6.77  & 1.1  & 12.8  & 100 & 800  &3.87&0.70\\
$10^4$ & $10^4$ & 200 & 143 & 13.53 & 0.5  & 6.4   & 100 & 1600  &5.14&0.93\\
$10^4$ & $10^4$ & 400 & 143 & 27.07 & 0.3  & 3.2   & 100 & 1600  &-  &-\\
$10^4$ & $10^5$ & 10  & 143 & 0.21  & 10.5 & 127.9 & 100 & 200   &2.32&0.42\\
$10^4$ & $10^5$ & 30  & 143 & 0.63  & 3.5  & 42.6  & 100 & 200   &3.09&0.56\\
$10^4$ & $10^5$ & 60  & 143 & 1.28  & 1.8  & 21.3  & 100 & 400   &3.37&0.61\\
$10^4$ & $10^5$ & 100 & 143 & 2.14  & 1.1  & 12.8  & 100 & 400   &3.42&0.62\\
$10^4$ & $10^5$ & 200 & 143 & 4.28  & 0.5  & 6.4   & 100 & 800  &4.31&0.78\\
$10^4$ & $10^5$ & 400 & 143 & 8.56  & 0.3  & 3.2   & 100 & 1600  &-  &-\\
$10^4$ & $10^6$ & 10  & 143 & 0.07  & 10.5 & 127.9 & 100 & 200   &0.94&0.17\\
$10^4$ & $10^6$ & 30  & 143 & 0.20  & 3.5  & 42.6  & 100 & 200   &0.88&0.16\\
$10^4$ & $10^6$ & 60  & 143 & 0.41  & 1.8  & 21.3  & 100 & 200   &0.94&0.17\\
$10^4$ & $10^6$ & 100 & 143 & 0.68  & 1.1  & 12.8  & 100 & 200   &1.10&0.20\\
$10^4$ & $10^6$ & 200 & 143 & 1.35  & 0.5  & 6.4   & 100 & 400  &1.10&0.20\\
$10^4$ & $10^6$ & 400 & 143 & 2.71  & 0.3  & 3.2   & 100 & 800  &1.10&0.20\\
$10^5$ & $10^4$ & 0  & 47  & -  & - & - & 140 & 140  &7.18&1\\
$10^5$ & $10^4$ & 10  & 47  & 0.68  & 18.9 & 129.3 & 140 & 280  &7.10&0.99\\
$10^5$ & $10^4$ & 30  & 47  & 2.03  & 6.3  & 43.1  & 140 & 280  &6.67&0.93\\
$10^5$ & $10^4$ & 60  & 47  & 4.06  & 3.1  & 21.5  & 140 & 560  &5.74&0.80\\
$10^5$ & $10^4$ & 100 & 47  & 6.77  & 1.9  & 12.9  & 140 & 1120  &4.74&0.66\\
$10^5$ & $10^4$ & 200 & 47  & 13.53 & 0.9  & 6.5   & 140 & 1120  &5.52&0.77\\
$10^5$ & $10^4$ & 400 & 47  & 27.07 & 0.5  & 3.2   & 140 & 1120  &-  &-\\
$10^5$ & $10^5$ & 10  & 47  & 0.21  & 18.9 & 129.3 & 140 & 280  &4.16&0.58\\
$10^5$ & $10^5$ & 30  & 47  & 0.63  & 6.3  & 43.1  & 140 & 280  &5.02&0.70\\
$10^5$ & $10^5$ & 60  & 47  & 1.28  & 3.1  & 21.5  & 140 & 560  &4.74&0.66\\
$10^5$ & $10^5$ & 100 & 47  & 2.14  & 1.9  & 12.9  & 140 & 1120  &4.52&0.63\\
$10^5$ & $10^5$ & 200 & 47  & 4.28  & 0.9  & 6.5   & 140 & 1120  &5.24&0.73\\
$10^5$ & $10^5$ & 400 & 47  & 8.56  & 0.5  & 3.2   & 140 & 1120  &-  &-\\
$10^5$ & $10^6$ & 10  & 47  & 0.07  & 18.9 & 129.3 & 140 & 280  &1.72&0.24\\
$10^5$ & $10^6$ & 30  & 47  & 0.20  & 6.3  & 43.1  & 140 & 280  &1.72&0.24\\
$10^5$ & $10^6$ & 60  & 47  & 0.41  & 3.1  & 21.5  & 140 & 280  &1.87&0.26\\
$10^5$ & $10^6$ & 100 & 47  & 0.68  & 1.9  & 12.9  & 140 & 280  &1.87&0.26\\
$10^5$ & $10^6$ & 200 & 47  & 1.35  & 0.9  & 6.5   & 140 & 560  &1.65&0.23\\
$10^5$ & $10^6$ & 400 & 47  & 2.71  & 0.5  & 3.2   & 140 & 1120 &1.29&0.18\\
\hline
\end{tabular}}
\tablefoot{Columns 1-3 show cloud mass, outflow temperature and velocity, followed by $\chi$ -- density contrast; $\mathcal{M}$ -- Mach number; cloud crushing ($t_{\rm cc}$) and acceleration ($t_{\rm acc}$) times; size of the simulated `box' in $y,z,x$; fragmentation and normalised fragmentation times $t_{\rm frag}$, $t_{\rm norm}$. Missing entries in the last two columns show non-fragmented systems.}

\end{table}

\FloatBarrier
\newpage

\section{Density maps}
\label{ap:density_maps}

Here we provide complementary (see Fig.~\ref{fig:EVO}) density maps for the $10^5$~M$_{\sun}$ cloud simulations with outflow temperatures of $10^4$, $10^5$,  and $10^6$~K. We briefly address the evolutionary differences from outflows of intermediate temperature (see Sect.~\ref{sec:Morphology}). 
In $10^4$~K outflows (Fig.~\ref{fig:EVO4K}), all but the lowest velocities result in supersonic interaction. The thermal pressure is the lowest of all simulations, and therefore compression by ram pressure is more pronounced. The dense outer shell of shocked gas does not form, increasing the cloud surface erosion and dispersal. In contrast, in $10^6$~K outflows, thermal pressure is dominant, and it rapidly compresses the cloud (Fig.~\ref{fig:EVO6K}). For the hot outflow case, only the last two rows are supersonic.

\begin{figure}[hbt!]
     \centering
         \includegraphics[width=0.7\textwidth]{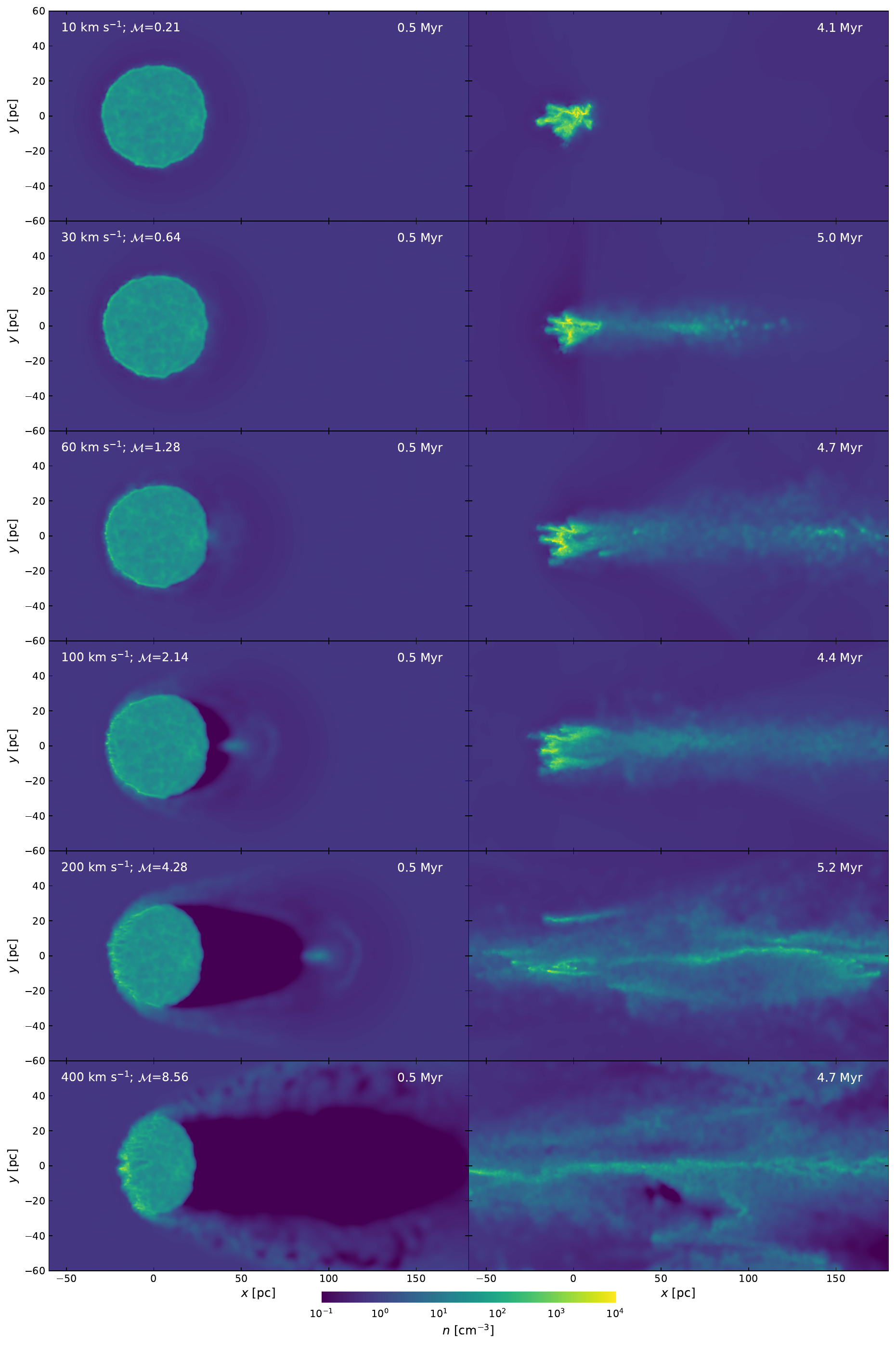}
        \caption{Particle density slices of a $10^5$~M$_{\sun}$ molecular cloud embedded in a $10^5$~K outflow. The left column shows the system's state after 0.5~Myr, the right one -- at $t_{\rm frag}$. Outflow velocity (Mach number), indicated at the top left of each panel, increase from 10~km~s$^{-1}\, (\mathcal{M}=0.21)$ in the top row to 400~km~s$^{-1}\, (\mathcal{M}=8.56)$ in the bottom one. Coordinates are centred on the cloud centre of mass, with outflow velocity in the positive $x$ direction.}
        \label{fig:EVO5K}
\end{figure}

\begin{figure}[hbt!]
     \centering
         \includegraphics[width=0.7\textwidth]{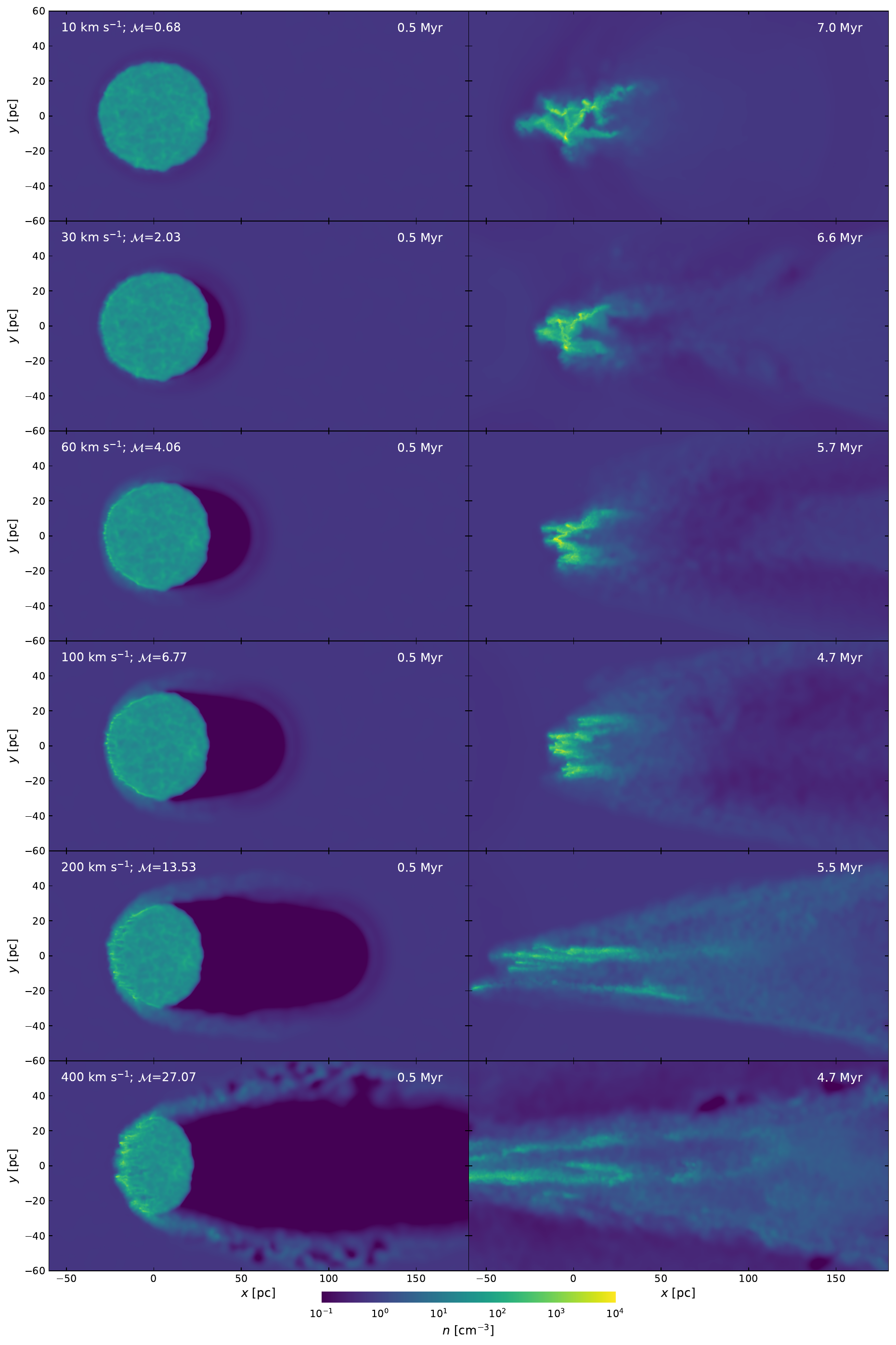}
        \caption{Same as Fig.~\ref{fig:EVO5K} but for $T_{\rm out}=10^4$~K.}
        \label{fig:EVO4K}
\end{figure}

\begin{figure}[hbt!]
     \centering
         \includegraphics[width=0.7\textwidth]{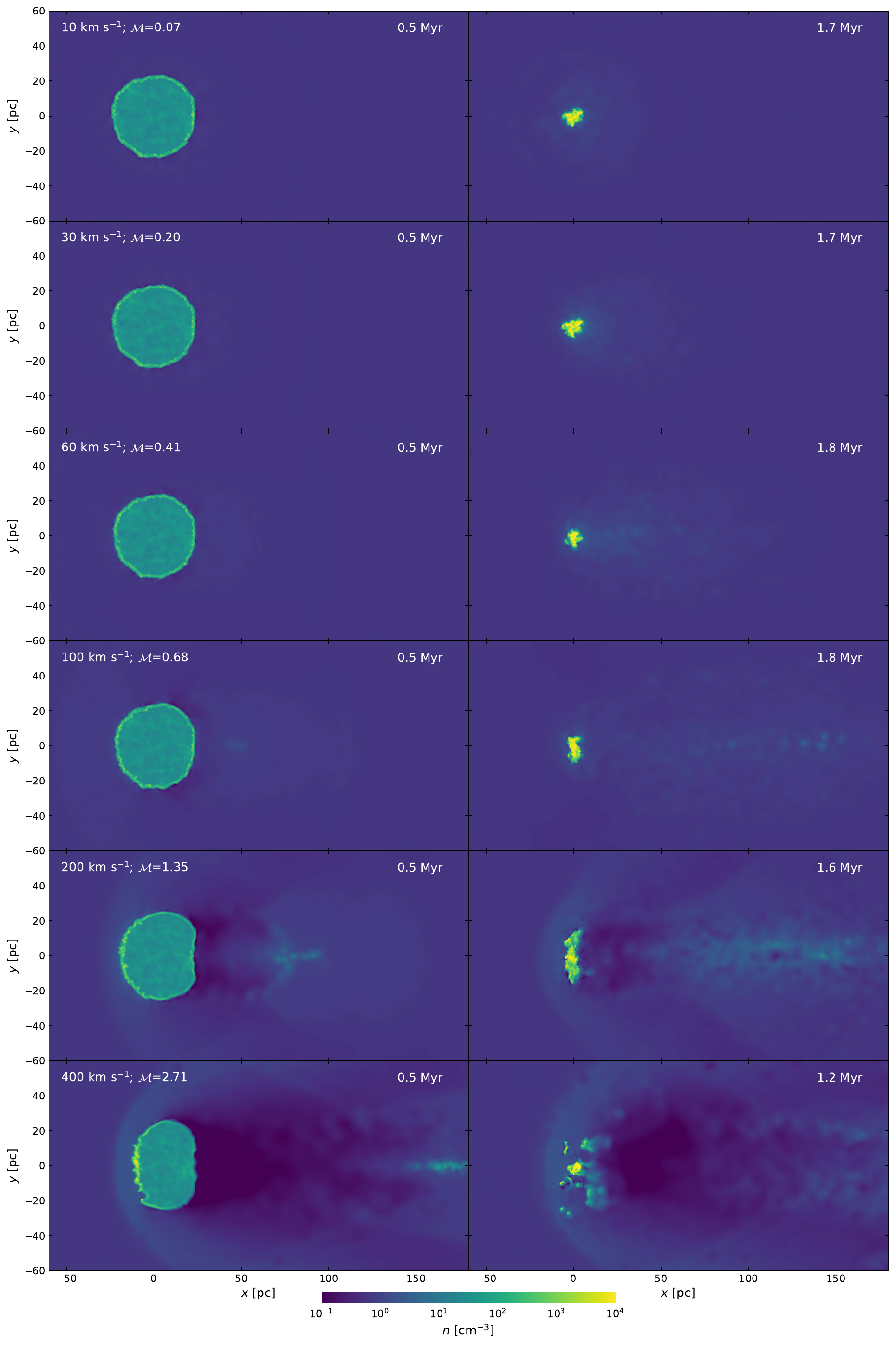}
        \caption{Same as Fig.~\ref{fig:EVO5K} but for $T_{\rm out}=10^6$~K.}
        \label{fig:EVO6K}
\end{figure}

\FloatBarrier
\newpage

\section{Gas state histograms}
\label{ap:Tn_diag}
\begin{figure}[hbt!]
     \centering
         \includegraphics[width=0.9\textwidth]{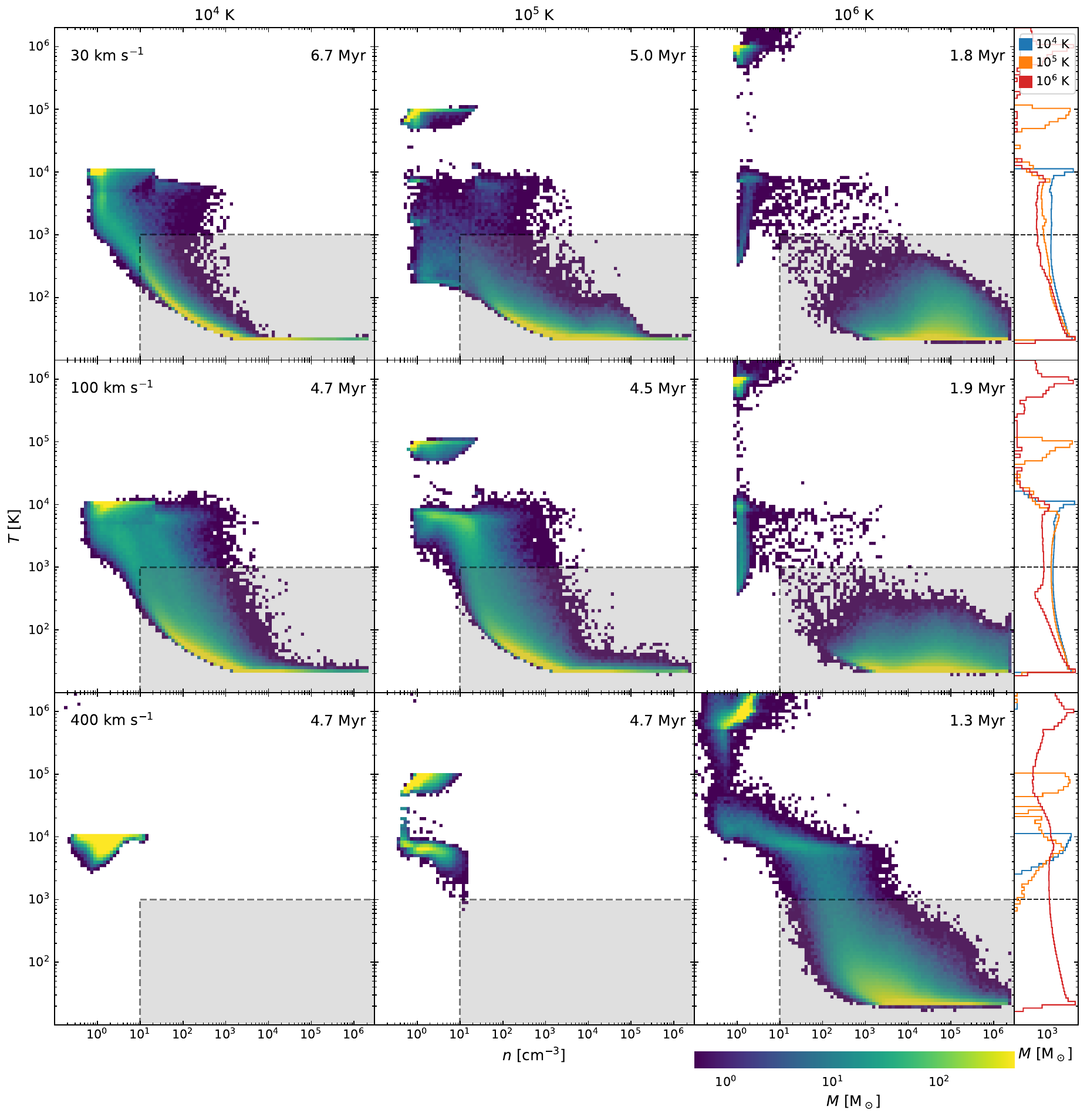}
        \caption{Temperature, number density histograms of $10^5$~M$_{\sun}$ clouds at fragmentation time (given in the upper right corner) with outflow temperatures of $10^4, 10^5, 10^6$~K (from left to right). Outflow velocity increases in rows - $30, 100, 400$~km~s$^{-1}$. The rightmost column shows the total gas mass in each temperature bin. The gas in the grey area is considered to comprise the molecular cloud.}
        \label{fig:Tn}
\end{figure}

In Sects.~\ref{sec:Cooling} and~\ref{sec:mass_cool} we introduce the selection criteria for gas to be considered part of the molecular cloud ($n>10$~cm$^{-3}$ and $T<1000$~K). In Fig.~\ref{fig:Tn} we show 2D histograms binned by temperature and density. The molecular cloud is composed of material contained within the grey box. We note that the precise choice of selection threshold values has little impact on the total mass of the cloud, since the majority of its mass has far more extreme values of both density and temperature.

\section{Evolution of cold gas}
\label{ap:cold_gas}
\begin{figure}
     \centering
         \includegraphics[width=0.9\textwidth]{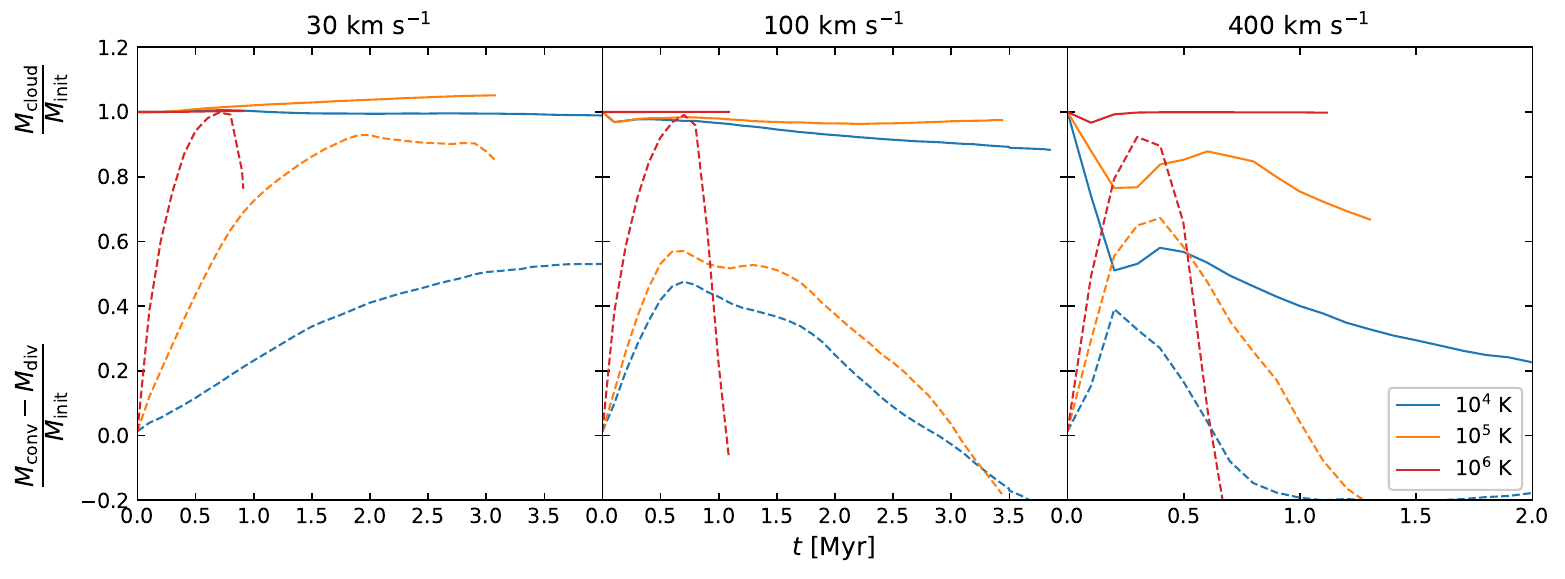}
        \caption{Evolution of cold gas mass. We present the ratio of cold gas mass to the initial cloud mass (solid line) and the ratio of the difference between converging and diverging cold gas mass (dashed line) to the initial cloud mass $\left(M_{\rm init}\right)$ for the $M_{\rm cl} = 10^4$~M$_{\sun}$ simulations. Outflow velocity increases from left to right, the colours indicate the outflow temperature, as given in the legend.}
        \label{fig:mass_4}
\end{figure}

\begin{figure}
     \centering
         \includegraphics[width=0.9\textwidth]{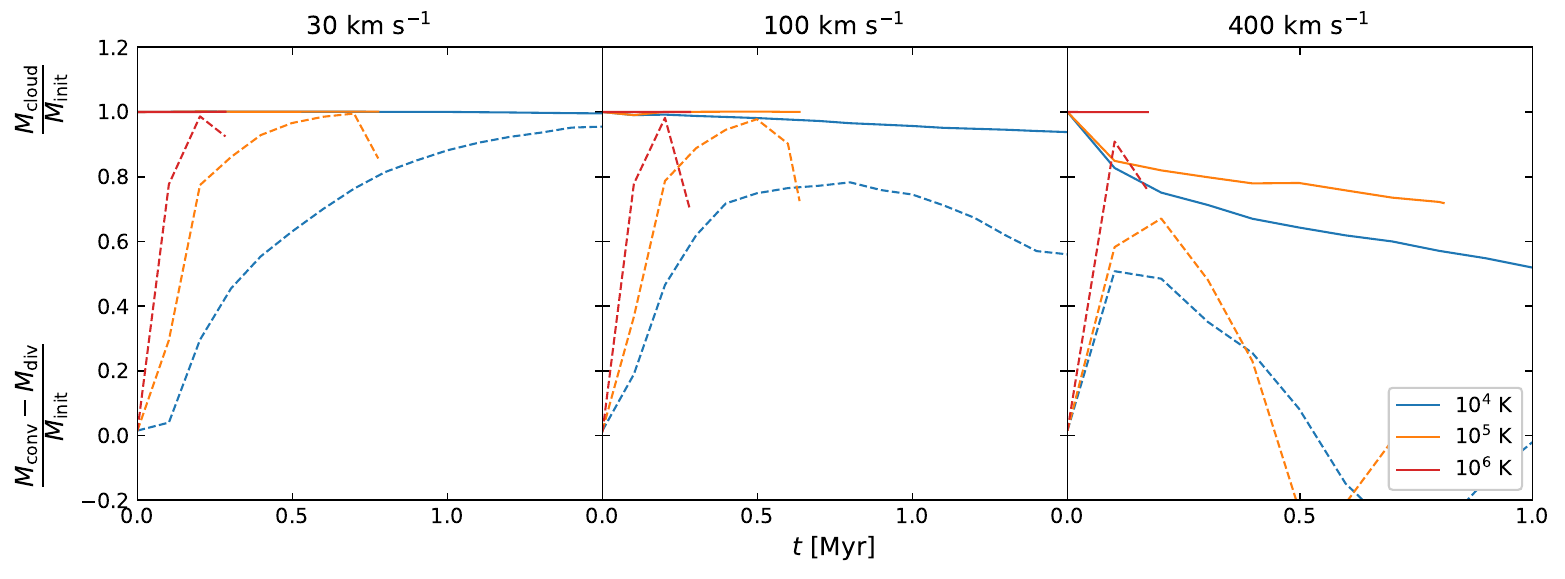}
        \caption{Same as in Fig.~\ref{fig:mass_4} but for $M_{\rm cl}=10^3$~M$_{\sun}$.}
        \label{fig:mass_3}
\end{figure}

In Sect. \ref{sec:mass_cool} we presented cold gas evolution for the $10^5$~M$_{\sun}$ cloud. Here we provide a brief overview for lower-mass clouds. The intermediate-mass cloud (Fig.~\ref{fig:mass_4}) evolution is similar to the $10^5$~M$_{\sun}$ cloud (see Fig.~\ref{fig:all_mass}) but with the main difference being shorter evolutionary times. Several models did not fragment (the ones that reach $4$~Myr in the rightmost panel). Despite not meeting fragmentation criteria, the clouds are not dispersed and retain a significant mass of cold gas. For the lowest mass clouds (Fig.~\ref{fig:mass_3}) the evolution times are even shorter, and in none of the models we observe the growth of cold gas mass.

\FloatBarrier

\FloatBarrier

\section{Cooling function}
\label{ap:cooling}

We used a cooling function as outlined by \citet{KakiuchiSuzuki:2024:}. We assumed low optical depth, a constant mean molecular weight and solar metallicity. The generic form of the cooling function is
\begin{equation}
    \rho_{\rm n} \mathcal{L}=n \left( -\Gamma+n \Lambda \right). 
    \label{eq:Heat_Cool}
\end{equation}
Here, $\rho_{\rm n}$ and $\mathcal{L}$ are the density and total loss rate of internal energy, while $n$ is the particle number density, and $\Lambda, \Gamma$ are volumetric cooling and heating rates, respectively.  The heating rate depends only on the temperature $T$ and cut-off temperature $T_{\rm cut}$:
\begin{equation}
\Gamma = 2 \times 10^{-26} \exp{\left ( - \frac{T}{T_{\rm cut}} \right)} \rm erg \ \rm s ^{-1}.
\label{eq:Heating_model}
\end{equation}
We choose $T_{\rm cut} = 5 \times10^4$~K. The cooling function is a piecewise combination of low- and high-temperature regions. At temperatures below $T < 10^4$~K, the cooling rate is
\begin{equation}
\Lambda_{\rm l} = 2 \times10^{-19} \ \exp{\left( \frac{-118400}{T+1000} \right)} + 2.8 \times 10^{-28} \sqrt{T}\ \exp{\left( \frac{-92}{T} \right) } \ {\rm erg} \ {\rm cm}^{-3} \ {\rm s}^{-1},
\label{eq:Cooling_model}
\end{equation}
and at higher temperatures,
\begin{equation}
\log_{10} \Lambda_{\rm h} = -156.919 + 84.2271(\log_{10} T) - 19.0317(\log_{10} T)^2 + 1.85211(\log_{10} T)^3 - 0.0658615(\log_{10} T)^4.
\end{equation}
To get the total cooling rate over the whole temperature range, we smoothly connected the cooling functions at a crossover point $T_{\rm b}$:
\begin{equation}
\Lambda=0.5 \left( \Lambda_{\rm l}(1-f) + \Lambda_{\rm h}(1+f) \right), \text {where} \  f=\tanh{\left( \frac{\log_{10} T - \log_{10} T_{\rm b}}{\log_{10} \Delta T_{\rm b}} \right)}.
\end{equation}
Function $f$ smooths the connection point between the two temperature regions. We set $T_{\rm b} = 10^4$~K and $\log_{10} \Delta T_{\rm b}=0.1$.

\end{appendix}

\end{document}